\documentclass{iopjournal}
\usepackage{xcolor}
\usepackage{float}
\usepackage{booktabs}
\expandafter\let\csname equation*\endcsname\relax

\expandafter\let\csname endequation*\endcsname\relax
\usepackage{varwidth}
\usepackage{amsmath}
\usepackage{amssymb}
\usepackage{bm}
\usepackage{graphicx}
\usepackage{wasysym}
\usepackage{url}

\newenvironment{cellvarwidth}[1][t]
    {\begin{varwidth}[#1]{\linewidth}}
      {\end{varwidth}}

\usepackage{array}
\usepackage{ragged2e}

\begin{document}
\articletype{Paper}

\title{Quest for quantum advantage: Monte Carlo wave-function simulations
of the CIM}
\author{Manushan Thenabadu, Run Yan Teh, Jia Wang, Simon Kiesewetter, Margaret
D Reid, Peter D Drummond}

\affil{Centre for Quantum Science and Technology Theory, Swinburne University
of Technology, Melbourne 3122, Australia}

\email{peterddrummond@protonmail.com}
\begin{abstract}
The Coherent Ising Machine (CIM) is a quantum network of optical parametric
oscillators (OPOs) intended to find ground states of the Ising model.
This is an NP-hard problem, related to several important minimization
problems, including the max-cut graph problem. In order to enhance
its potential performance, we analyze the coherent coupling strategy
for the CIM in a highly quantum regime. To explore this limit, without
assuming gaussianity, we employ accurate numerical simulations. Due
to the inherent complexity of the system, the maximum network size
is limited. While master equation methods can be used, their scalability
diminishes rapidly for larger systems. Instead, we use Monte Carlo
wave-function methods, which scale as the wave-function dimension,
and use large numbers of samples. These simulations involve Hilbert
spaces exceeding $10^{7}$ dimensions. To evaluate success probabilities,
we use quadrature probabilities. We demonstrate the potential for
quantum computational advantage by reducing the time required to reach
maximum success probability in a low-dissipation regime enabled by
initial quantum superpositions and entanglement. Furthermore, we
demonstrate that tailored time-dependent couplings can amplify these quantum effects. Comparisons with
classical CIM models give evidence that quantum tunneling effects
in this strong coupling limit can overcome trapping in false minima.
This can greatly increase success rates, indicating a potential for
quantum advantage. Finally, we perform a coherence analysis based
on the state purity to examine the role of quantum coherence in CIM
performance and to determine how state purity correlates with improved
optimization outcomes.
\end{abstract}

\section{Introduction}

Quantum computational advantage has been an important motivation for
quantum computing researchers \cite{Feynman1982simulating,nielsen2010quantum}.
The goal is to build a quantum device that solves problems faster
than digital hardware. There are many recent experiments motivated
by this goal \cite{arute2019quantum,zhong2020quantum,wu2021strong,zhu2022quantum,madsen2022quantum}.
However, the resulting data errors raise the issue that the device
may not solve the original problem \cite{drummond2022simulating,MartinezCifuentes2023classicalmodelsmay,dellios2025validation}.
Additionally, a quantum computer with broadly useful applications
is still not realized \cite{daley2022practical}. Here we study quantum
advantage in practical optimization problems. We use the coherent
Ising machine (CIM) as the computing architecture, since it targets
such problems.

The CIM is intended to solve the Ising model ground state, initially
developed to model phase transitions in magnetic materials \cite{Newell1953theory}.
The Ising model is a network of discrete spins $\sigma_{k}=\pm1$
with nearest-neighbor interactions. More generally, we take into account
arbitrary couplings $J_{ij}$ and an external magnetic field $h_{j}$,
at each site. The energy function of the model has the form, 
\begin{equation}
E(\bm{\sigma})=-\underset{i,j}{\sum}J_{ij}\sigma_{i}\sigma_{j}-\underset{j}{\sum}h_{j}\sigma_{j}.
\end{equation}
This has many applications to optimization problems across fields
including computer science, medicine, logistics, finance, telecommunication
and machine learning \cite{applications_1,crescenzi1997approximation,sankar2024benchmarking}.
On mapping NP-hard and NP-complete problems to the Ising Hamiltonian
\cite{lucas2014ising,glover2022quantum}, one finds that the ground state
corresponds to the optimal solution for these problems. Conventional
computers face significant challenges in solving such complex optimization
problems due to their exponential time complexity. In our study, the
Ising model is mapped onto a non-equilibrium quantum optical system,
namely the coherent Ising Machine (CIM). We investigate quantum dynamics
in deeply quantum regimes, to determine an optimal strategy that leads
to a ground state that can be measured.

This computing architecture is inspired by the Ising model. It attempts
to use quantum mechanics to potentially overcome classical computational
limitations, although it is not a universal, gate-based architecture.
The CIM model comprises a network of degenerate optical parametric
oscillators (DOPOs), where each DOPO element corresponds to an Ising
spin. Coherent coupling between the DOPO elements is achieved using
an optical delay line scheme \cite{Yamamoto2017}. In this scheme,
DOPO pulses circulate within an optical ring cavity, with each pulse
representing an Ising spin. The phase of the pulse determines whether
it is a spin-up or spin-down state.

To implement the $J_{ij}$ coupling between DOPO pulses, one strategy
is to pick off a portion of each pulse in the optical ring and feed
it through an optical phase-sensitive amplifier (PSA), followed by
delay lines equipped with intensity and phase modulators, thus allowing
for arbitrary coupling between any two pulses within the system, determined
by the coupling coefficient $J_{ij}$. A small-scale network
was first experimentally realized by Marandi et al. \cite{Marandi2014},
using a time-division multiplexing scheme within a single ring resonator,
with spin-spin interactions implemented through mutual injections
of pulses using delay interferometers. {This was extended to $16$ \cite{Takata:2016aa} and later to $10^{4}$ DOPOs \cite{inagaki2016large}, which is the largest network in the optical delay line scheme to date.} Since then, large-scale
networks comprising up to $10^{5}$ DOPOs \cite{McMahon2016,Takesue2025,hamerly2019experimental,honjo2021100}
have been investigated, although these use an incoherent measurement
feedback (MFB) strategy which reduces quantum effects. Other proposals
suggest different physical implementations, such as opto-mechanics
\cite{Rah2023} or the superconducting Kerr effect \cite{Goto_SciRep2016,goto_CIM}. { Many-body interactions can also be included in an Ising machine. An experiment implementing four-body interactions in an Ising machine using measurement feedback has been reported \cite{kumar_2020}. The inclusion of many-body interactions further expands the computationally hard problem class that can be solved by an Ising machine. All experiments to date have weak nonlinearity that masks the effects of quantum fluctuations and superpositions.}
Here we return to the original feedback scheme, but investigate the
deep quantum regime in which quantum fluctuations and superpositions
have large effects, which also could be applicable to other nonlinear
quantum technologies.  

To simulate the quantum dynamics of such parametric systems, master
equation methods can be used \cite{Teh2020,yamamura2017quantum}, but these scale quadratically
in the Hilbert space dimension, making them impractical for larger
systems with many coupled oscillators. Positive-P phase-space methods
\cite{Kiesewetter2022,Drummond_generalizedP1980,shoji2017quantum} and truncated Wigner methods \cite{maruo2016truncated}, which are useful for
the present experimental parameters, and are scalable, have
drawbacks in highly quantum regimes due to sampling errors \cite{Deuar_RPA2002}. A third
approach used for MFB simulations \cite{Ng2022,single_photon_yoshi},
the gaussian approximation, is inapplicable when there are non-gaussian
quantum superpositions {(for a review on gaussianity of a quantum state, see for example: \cite{Walschaers_PRX2021,Yanagimoto:24,Jankowski:24})}. { When the states of the system are gaussian states \cite{drummond2022simulating,Ng2022,serafini2004minimum,olsen2009numerical}, they can be treated as classical states with fluctuations. Also, a gaussian state under gaussian operations, such as displacement and squeezing of states, remains a gaussian state. Hence, classical Wigner simulations can track the evolution of these states quite accurately. This fact is captured formally by the continuous variable version of the Gottesman-Knill theorem \cite{Bartlett_2002}. However, the nonlinear parametric interaction in a CIM causes a state to deviate from a gaussian state, especially as we use strong nonlinear parametric interactions.} In this study, we employ Monte Carlo wave-function
(MCWF) methods \cite{molmer1993monte}, which scale linearly in the
Hilbert space dimension, and can treat any quantum state. We report
on simulations that exceed $10^{7}$ Hilbert dimensions. Although
our simulations are still in the small mode limit of $M<6$, they
are beyond the reach of master equations, and can treat extreme quantum
regimes.

We investigate several optimization problems, ranging from anti-ferromagnetic
ground states to the max-cut graph problem. We evaluate the success
probability through $x$-quadrature measurements of the DOPO outputs.
The success probability is calculated from the joint probabilities
of quadratures $P(x_{1,}x_{2},...,x_{M})$, with $x_{i}\ge0$ as a
positive spin and $x_{i}<0$ as a negative spin. We investigate multiple
strategies for the initial quantum state, while dynamically changing the
coupling strength and other parameters, to improve the success probability
of the simulations. We find a speed up in performance and improved
success probabilities with non-classical initial states, using strategies
in which the system transitions from an initial highly quantum regime
with quantum superpositions, to a final classical regime where the
resulting configuration is read out. Comparisons with classical mean-field
solutions give evidence that there is a potential quantum advantage
in cases where classical trapping in false minima limit the performance
of classical strategies.

Because an accurate simulation of this type of computer is harder
than the Ising problem, our conclusions are indicative. The small-scale
devices that we simulate cannot solve large Ising problems. Nevertheless,
we can investigate whether quantum superpositions have a potential
for quantum advantage. We find evidence for more rapid convergence
to a solution when quantum superpositions are present. We also show
that a limitation of this approach is decoherence due to absorption
in the spin couplings. This indicates that other, more coherent coupling
methods may be even more beneficial.

\section{Hamiltonian and Master Equation\label{sec:Hamiltonian-and-Master}}

{Here, we  analyze the use of the Coherent Ising Machine (CIM) to solve combinatorial optimization
problems in highly quantum regimes. Although algorithms exist that
can solve the Ising model with more than 100 spins \cite{SBM}, conventional
computers struggle with larger numbers of spins due to their exponential
time complexity. A CIM model with a quantum architecture may offer
advantages in performance, speed, and energy efficiency compared to
its classical counterpart \cite{q_annealing}. In this study, we
evaluate the system dynamics by utilising the full Hilbert space,
except for a number cut-off. Since this is a dissipative problem, a master equation is required to treat it. 
However, for improved scaling properties, it is advantageous to use stochastic wave-function methods,
rather than direct solutions of the master equation. 
For system evolution, both jump and continuous
methods exist \cite{Haribara2016,SSE,SSE_General} as another alternative.
Here, we employ the Monte Carlo Wave Function (MCWF) method in our
numerical simulations of the CIM to assess the system dynamics, and
we review the procedure in detail in Appendix A, but
first we give a detailed explanation of the master equation derivation.}

{Although the experiments use propagating pulses, we simplify this as in other theoretical studies by assuming that each pulse is effectively a single mode, since while one can treat propagation, this substantially increases the problem complexity.} We begin by considering a single DOPO, which can be approximately
mapped when above threshold to an individual Ising spin with two possible
configurations---spin-up or spin-down. It is described by the Hamiltonian
\cite{drummond1980non,teh2020dynamics,Drummond_OpActa1981}: 
\begin{align}
\mathcal{H} & =\omega_{s}a^{\dagger}a+a^{\dagger}\Gamma_{s}+a\Gamma_{s}^{\dagger}+\omega_{p}a_{p}^{\dagger}a_{p}+a_{p}^{\dagger}\Gamma_{p}+a_{p}\Gamma_{p}^{\dagger}\nonumber \\
 & +\frac{i\kappa}{2}\left(a_{p}a^{\dagger2}-a_{p}^{\dagger}a^{2}\right)+i\epsilon\left(a_{p}^{\dagger}e^{-i\omega_{p}t}-a_{p}e^{i\omega_{p}t}\right).
\end{align}

Here $a_{p}$ is the pump mode, $a_{}$ is the signal/idler mode,
$\omega_{p}$ and $\omega_{s}$ are the pump and signal mode frequencies,
and $\kappa$ is the coupling strength between the optical cavity
modes, in units such that $\hbar=1$. The first term represents the
free evolution and linear damping of the optical cavity modes, the
second term is the coupling between the pump and signal modes, and
the third term represents the driving of the pump mode by an external
field of amplitude $\epsilon$ with frequency $\omega_{p}$. After
adding amplitude damping to both modes at rates $\gamma$ and $\gamma_{p}$
respectively, we must use a master equation or similar dissipative
evolution equation to describe the system.

\subsection{Pump adiabatic elimination}

 When the pump mode decay rate is much larger than the signal mode
decay rate ($\gamma_{p}>>\gamma$), the pump mode can be removed through
adiabatic elimination \cite{Kinsler_PRA1991,carmichael2009statistical}.
The simpler Hamiltonian that results in the rotating frame is ,
\begin{align}
\mathcal{H}_{s} & =\bar{\Delta}a^{\dagger}a+i\gamma\frac{\lambda}{2}\left(a^{\dagger2}-a^{2}\right)\nonumber \\
 & +\left(a^{\dagger}\Gamma_{s}+a\Gamma_{s}^{\dagger}\right)+\frac{\gamma g^{2}}{2\gamma_{p}}\left(a^{2}\Gamma_{p}^{\dagger}+a^{\dagger2}\Gamma_{p}\right),\label{eq:Hamiltonian}
\end{align}
where $a$ is the signal mode and $\Gamma_{s}$, $\Gamma_{p}$
are the reservoir operators. Here $\Delta=\omega_{s}-\omega_{p}/2$
is the detuning and $\lambda$ the pump driving strength . This allows
us to define $g$ as a dimensionless two-photon dissipation rate and
$\lambda$ as a single photon driving rate, where:
\begin{align}
\lambda & =\frac{|\kappa\epsilon|}{\gamma\gamma_{p}}\nonumber \\
g & =\sqrt{\frac{|\kappa|^{2}}{2\gamma\gamma_{p}}}.
\end{align}

We define a characteristic coherent amplitude of 
\begin{equation}
\alpha_{c}=\frac{\sqrt{\lambda}}{g}.\label{eq:cat-amplitude}
\end{equation}
The steady state is known exactly, and when above threhold \cite{drummond1980non}
can be understood approximately as a mixture of coherent states $|\pm\alpha_{c}\rangle$
\cite{Krippner_PRA1994} . This gives a characteristic photon number
scale of $n_{c}=|\alpha_{c}|^{2}$. Prior to this, the dynamical behavior
in highly quantum regimes with $g\gtrsim1$ \cite{Krippner_PRA1994,Wolinsky_PRL1988,Gilles_PRA1994}
leads to an approximate transient Schr{\"o}dinger cat state: 
\begin{equation}
|\Psi_{C}\rangle=N_{C}\left[|\alpha_{c}\rangle+|-\alpha_{c}\rangle\right],
\end{equation}
which decays to the steady state from decoherence caused by the single
photon decays \cite{Munro_PRA1995}, giving an exact analytic solution
\cite{drummond1980non}.

\subsection{Multimode dynamical master equation}

We now derive the full master equation, using a coupled cavity model
with coherent feedback injection \cite{Marandi_CIM_Nature2014,goto_CIM,Takata2015}.
The degenerate optical parametric oscillators (DOPOs) are mutually
coupled via injection paths, and our goal is to adiabatically eliminate
the coupling modes to obtain a reduced description involving only
the DOPO signal modes. The complete Hamiltonian of the coherent Ising
machine (CIM) can be written as \cite{Takata2015},
\begin{align}
\mathcal{H} & =\mathcal{H}_{free}+\mathcal{H}_{int}+\mathcal{H}_{pump}+\mathcal{H}_{res}+\mathcal{H}_{BS}
\end{align}
where each term is defined as follows:

\paragraph{Free evolution:}

The uncoupled dynamics of the pump and signal modes of each DOPO,
together with the signal modes of the injection paths are described
by
\begin{align}
\mathcal{H}_{free} & =\underset{j=1}{\overset{M}{\sum}}\left(\omega_{p}{a}_{p,j}^{\dagger}{a}_{p,j}+\hbar\omega_{s}{a}_{j}^{\dagger}{a}_{j}\right)\nonumber \\
 & +\underset{1\le i<j}{\overset{M}{\sum}}\omega_{s}{b}_{ij}^{\dagger}{b}_{ij}.
\end{align}

Here $M$ is the number of DOPO's, and the number of injection paths
is $\frac{1}{2}(M^{2}-M)$. Operators $({a}_{p,j},{a}_{p,j}^{\dagger})$
and $({a}_{j},{a}_{j}^{\dagger})$ annihilate and create photons
in the pump and signal modes of the $j^{th}$ DOPO respectively, while
$({b}_{ij},{b}_{ij}^{\dagger})$ act on the injection path
that couples the $i$ and $j$ modes. The pump frequency is tuned
to satisfy $\omega_{p}=2\omega_{s}$, so $\bar{\Delta}=0$.

\paragraph{Parametric interaction:}

Photon conversion between the pump and signal modes is governed by
\begin{align}
\mathcal{H}_{int} & =i\underset{j=1}{\overset{M}{\sum}}\frac{\kappa}{2}\left({a}_{j}^{\dagger2}{a}_{p,j}-{a}_{p,j}^{\dagger}{a}_{j}^{2}\right),
\end{align}
where $\kappa$ is the quadratic nonlinear coupling coefficient. Each
pump photon lost (gained) produces two signal photons gained (lost).

\paragraph{External driving:}

The cavity is pumped by a classical source of strength $\epsilon$,
\begin{align}
\mathcal{H}_{pump} & =i\epsilon\underset{j=1}{\overset{M}{\sum}}\left({a}_{p,j}^{\dagger}e^{-i\omega_{p}t}-\epsilon_{p}{a}_{p,j}e^{i\omega_{p}t}\right).
\end{align}

\paragraph{System--reservoir coupling:}

Dissipation of the DOPO pump and signal modes and of the injection
modes is described by
\begin{align}
\mathcal{H}_{res} & =\underset{j=1}{\overset{M}{\sum}}\left({a}_{j}\hat{\Gamma}_{s,j}^{\dagger}+\hat{\Gamma}_{s,j}{a}_{j}^{\dagger}+{a}_{p,j}\hat{\Gamma}_{p,j}^{\dagger}+\hat{\Gamma}_{p,j}{a}_{p,j}^{\dagger}\right)\nonumber \\
 & +\underset{1\le i<j}{\overset{M}{\sum}}\left({b}_{ij}\hat{\Gamma}_{c,ij}^{\dagger}+\hat{\Gamma}_{c,ij}{b}_{ij}^{\dagger}\right),
\end{align}
where the $\hat{\varGamma}$ operators represent the bath modes.

\paragraph{Beam-splitter coupling:}

The interaction between DOPO signal modes and injection-path modes
is
\begin{align}
\mathcal{H}_{BS} & =i\underset{1\le i<j}{\overset{M}{\sum}}\zeta_{ij}\left({b}_{ij}{a}_{i}^{\dagger}-{b}_{ij}^{\dagger}{a}_{i}+\right.\nonumber \\
 & \left.+\frac{J_{ij}}{|J_{ij}|}{a}_{j}{b}_{ij}^{\dagger}-\frac{J_{ij}}{|J_{ij}|}{a}_{j}^{\dagger}{b}_{ij}\right),
\end{align}
where $\zeta_{ij}$ is the beam-splitter coupling coefficient between
$i^{th}$ and $j^{th}$ DOPO's, and $\frac{J_{ij}}{|J_{ij}|}=\pm1$
sets the relative phase on each path, with all coefficients
assumed real.

\subsection{Elimination of injection modes}

Following standard adiabatic elimination of the pump modes \cite{carmichael_book,carmichael2009statistical},
in the interaction picture, we define $\rho_{c}(t)$ as the density
matrix of the combined system of parametric down-converted modes plus
injection modes. Transforming $\rho_{c}(t)=e^{-i\omega_{s}\left(\sum_{i}a_{i}^{\dagger}a_{i}+\sum_{i<j}b_{ij}^{\dagger}b_{ij}\right)t}\tilde{\rho}e^{i\omega_{s}\left(\sum_{i}a_{i}^{\dagger}a_{i}+\sum_{i<j}b_{ij}^{\dagger}b_{ij}\right)t}$,
which removes linear frequency terms, the master equation for the
combined system including the injection modes in the Born-Markov approximation
is: 
\begin{align}
\dot{\tilde{\rho}} & =\left(\mathcal{L}_{s}+\mathcal{L}_{c}+\mathcal{L}_{sc}\right)\tilde{\rho},
\end{align}
with Liouvillians
\begin{align}
\mathcal{L}_{s}\tilde{\rho} & =\overset{M}{\underset{j=1}{\sum}}\frac{\gamma\lambda}{2}\left[a_{j}^{\dagger2}-a_{j}^{2},\tilde{\rho}\right]+\nonumber \\
 & +\overset{M}{\underset{j=1}{\sum}}\gamma\left[2a_{j}\tilde{\rho}a_{j}^{\dagger}-a_{j}^{\dagger}a_{j}\tilde{\rho}-\tilde{\rho}a_{j}^{\dagger}a_{j}\right]\nonumber \\
 & +\overset{M}{\underset{j=1}{\sum}}\frac{\gamma g^{2}}{2}\left[2\left(a_{j}^{2}\tilde{\rho}a_{j}^{\dagger2}\right)-\left(a_{j}^{\dagger2}a_{j}^{2}\tilde{\rho}\right)-\left(\tilde{\rho}a_{j}^{\dagger2}a_{j}^{2}\right)\right],
\end{align}
\begin{align}
\mathcal{L}_{c}\tilde{\rho} & =\underset{1\le i<j}{\overset{M}{\sum}}\gamma_{c}\left[2b_{ij}\tilde{\rho}b_{ij}^{\dagger}-b_{ij}^{\dagger}b_{ij}\tilde{\rho}-\tilde{\rho}b_{ij}^{\dagger}b_{ij}\right],
\end{align}
and
\begin{align}
\mathcal{L}_{sc}\tilde{\rho} & =-i\left[H_{BS},\tilde{\rho}\right]\nonumber \\
 & =\underset{1\le i<j}{\overset{M}{\sum}}\zeta_{ij}\left[b_{ij}a_{i}^{\dagger}-b_{ij}^{\dagger}a_{i}+\frac{J_{ij}}{|J_{ij}|}a_{j}b_{ij}^{\dagger}-\frac{J_{ij}}{|J_{ij}|}a_{j}^{\dagger}b_{ij},\tilde{\rho}\right].
\end{align}

Here $a_{i}$ and $a_{i}^{\dagger}$ are the annihilation and creation
operators for the signal modes, $b_{ij}$ and $b_{ij}^{\dagger}$
are the annihilation and creation operators for the injection modes,
$\gamma$ and $\gamma_{c}$ are the single-photon decay rates of the
DOPO and injection modes, respectively, while $\lambda$ and $g$
are the pump and nonlinear coefficients defined in Sec. \ref{sec:Hamiltonian-and-Master}.

To obtain a master equation involving only the DOPO signal modes,
we treat the injection modes as fast variables and trace them out
using super-operator algebra.  We isolate the interaction between
the two modes by the transformation,
\begin{align}
\bar{\rho}(t) & =e^{-\left(\mathcal{L}_{s}+\mathcal{L}_{c}\right)t}\tilde{\rho}(t).
\end{align}
This gives a picture in which the transformed density matrix $\bar{\rho}$
only evolves due to coupling with the injection modes: 
\begin{align}
\dot{\bar{\rho}}(t) & =\bar{\mathcal{L}}_{sc}(t)\bar{\rho},
\end{align}
where
\begin{align}
\bar{\mathcal{L}}_{sc}(t) & =e^{-\left(\mathcal{L}_{s}+\mathcal{L}_{c}\right)t}\mathcal{L}_{sc}e^{\left(\mathcal{L}_{s}+\mathcal{L}_{c}\right)t}.
\end{align}

The reduced density operator in this type of interaction picture is
obtained by tracing over the injection modes,
\begin{align}
\bar{\sigma}(t) & =\mathrm{tr}_{c}\left[\bar{\rho}(t)\right],
\end{align}
giving the equation of motion,
\begin{align}
\dot{\bar{\sigma}} & =\mathrm{tr}_{p}\left[\bar{\mathcal{L}}_{sp}(t)\rho(0)\right]+\int_{0}^{t}\mathrm{tr}_{p}\left[\bar{\mathcal{L}}_{sp}(t)\bar{\mathcal{L}}_{sp}(t')\bar{\rho}(t')\right]dt'.
\end{align}

The first term $\mathrm{tr}_{c}\left[\bar{\mathcal{L}}_{sc}(t)\rho(0)\right]=0$
by assuming $\rho(0)\approx\sigma(t)\left(|0\rangle\langle0|\right)_{c}$,
hence the equation of motion simplifies to
\begin{align*}
\dot{\bar{\sigma}} & =\int_{0}^{t}\mathrm{tr}_{c}\left[\bar{\mathcal{L}}_{sc}(t)\bar{\mathcal{L}}_{sc}(t')\bar{\sigma}(t')\left(|0\rangle\langle0|\right)_{c}\right]dt'.
\end{align*}
We can now write the super-operator $\bar{\mathcal{L}}_{sc}$ as
\begin{align}
\bar{\mathcal{L}}_{sc}(t) & =\underset{1\le i<j}{\overset{}{\sum}}\zeta_{ij}\left\{ \bar{S}_{i}(t)\bar{P}_{i}(t)-\bar{S}_{j}(t)\bar{P}_{j}(t)\right.-\nonumber \\
 & \left.-\bar{S}_{j}^{\dagger}(t)\bar{P}_{j}^{\dagger}(t)+\bar{S}_{i}^{\dagger}(t)\bar{P}_{i}^{\dagger}(t)\right\} ,
\end{align}
where the operators $\bar{S}_{\mu}(t)$ and $\bar{P}_{\mu}(t)$ encode
the time evolution of the DOPO and injection degrees of freedom, given
as:
\begin{align}
\bar{S}_{i}(t)\tilde{\rho} & =e^{-\mathcal{L}_{s}t}\left\{ \left(a_{i}^{\dagger}\tilde{\rho}\right)-\frac{J_{ij}}{|J_{ij}|}\left(a_{j}^{\dagger}\tilde{\rho}\right)\right\} e^{\mathcal{L}_{s}t}\nonumber \\
\bar{S}_{j}(t)\tilde{\rho} & =e^{-\mathcal{L}_{s}t}\left\{ \left(a_{i}\tilde{\rho}\right)-\frac{J_{ij}}{|J_{ij}|}\left(a_{j}\tilde{\rho}\right)\right\} e^{\mathcal{L}_{s}t}\nonumber \\
\bar{P}_{i}(t)\tilde{\rho} & =e^{-\gamma_{c}t}\left(b_{i}\tilde{\rho}\right)\nonumber \\
\bar{P}_{j}(t)\tilde{\rho} & =e^{\gamma_{c}t}\left(b_{i}^{\dagger}\tilde{\rho}\right)+\left(e^{-\gamma_{c}t}-e^{\gamma_{c}t}\right)\left(\tilde{\rho}b_{i}^{\dagger}\right)\nonumber \\
\bar{P}_{i}^{\dagger}(t)\tilde{\rho} & =e^{-\gamma_{c}t}\left(\tilde{\rho}b_{i}^{\dagger}\right)\nonumber \\
\bar{P}_{j}^{\dagger}(t)\tilde{\rho} & =e^{\gamma_{c}t}\left(\tilde{\rho}b_{i}\right)+\left(e^{-\gamma_{c}t}-e^{\gamma_{c}t}\right)\left(b_{i}\tilde{\rho}\right).
\end{align}

{}Evaluating the second-order term and
defining the operator $L_{i,j}=a_{i}-\frac{J_{ij}}{|J_{ij}|}a_{j}$,
we obtain
\begin{align}
\dot{\bar{\sigma}} & =\underset{1\le i<j}{\overset{M}{\sum}}\frac{\zeta_{ij}^{2}}{\gamma_{c}}e^{-\mathcal{L}_{s}t}\left[2\left(L_{i,j}e^{\mathcal{L}_{s}t}\bar{\sigma}(t)L_{i,j}^{\dagger}\right)-\left(L_{i,j}^{\dagger}L_{i,j}e^{\mathcal{L}_{s}t}\bar{\sigma}(t)\right)\right.\nonumber \\
 & \,\left.-\left(e^{\mathcal{L}_{s}t}\bar{\sigma}(t)L_{i,j}^{\dagger}L_{i,j}\right)\right].
\end{align}
{Finally, we invert the transformation to
get the equation of motion for the reduced density matrix $\rho=e^{\mathcal{L}_{s}t}\bar{\sigma}$}
,
\begin{align}
\dot{\rho} & =\mathcal{L}_{s}e^{\mathcal{L}_{s}t}\bar{\sigma}+e^{\mathcal{L}_{s}t}\dot{\bar{\sigma}},
\end{align}
which leads to{
\begin{align}
\dot{\rho} & =\mathcal{L}_{s}\rho+\underset{1\le i<j}{\overset{M}{\sum}}\frac{\zeta_{ij}^{2}}{\gamma_{c}}\left[2L_{i,j}\rho L_{i,j}^{\dagger}-L_{i,j}^{\dagger}L_{i,j}\rho-\rho L_{i,j}^{\dagger}L_{i,j}\right]\nonumber \\
 & =\overset{M}{\underset{j=1}{\sum}}\frac{\gamma\lambda}{2}\left[a_{j}^{\dagger2}-a_{j}^{2},\rho\right]+\overset{M}{\underset{j=1}{\sum}}\gamma\left[2a_{j}\rho a_{j}^{\dagger}-a_{j}^{\dagger}a_{j}\rho-\rho a_{j}^{\dagger}a_{j}\right]\nonumber \\
 & +\overset{M}{\underset{j=1}{\sum}}\frac{\gamma g^{2}}{2}\left[2a_{j}^{2}\rho a_{j}^{\dagger2}-a_{j}^{\dagger2}a_{j}^{2}\rho-\rho a_{j}^{\dagger2}a_{j}^{2}\right]\nonumber \\
 & +\underset{1\le i<j}{\overset{M}{\sum}}\frac{\zeta_{ij}^{2}}{\gamma_{c}}\left[2L_{i,j}\rho L_{i,j}^{\dagger}-L_{i,j}^{\dagger}L_{i,j}\rho-\rho L_{i,j}^{\dagger}L_{i,j}\right].
\end{align}
}

\subsection{Dimensionless master equation}

We now change to a dimensionless time, $\tau=t\gamma$, scaled with
respect to the single-photon decay rate $\gamma$. The following
master equation is applicable to a CIM of $M$ coherently coupled
oscillators, as introduced in previous work \cite{Marandi_CIM_Nature2014,goto_CIM,Takata2015},
where our model includes time-dependent pumping $\lambda(\tau)$ and
a dynamical two-photon decay $g\left(\tau\right)$, so that:
\begin{align}
\frac{\partial\rho}{\partial\tau} & =\underset{i}{\sum}\left\{ \frac{\lambda(\tau)}{2}\left[a_{i}^{\dagger}a_{i}^{\dagger}-a_{i}a_{i},\rho\right]\right\} \nonumber \\
 & +\underset{i}{\sum}\left\{ 2a_{i}\rho a_{i}^{\dagger}-a_{i}^{\dagger}a_{i}\rho-\rho a_{i}^{\dagger}a_{i}\right\} \nonumber \\
 & +\underset{i}{\sum}\left\{ \frac{g(\tau)^{2}}{2}\left(2a_{i}^{2}\rho a_{i}^{\dagger2}-a_{i}^{\dagger2}a_{i}^{2}\rho-\rho a_{i}^{\dagger2}a_{i}^{2}\right)\right\} \nonumber \\
 & +\sum_{i<j}\xi_{0}\left(\tau\right)|J_{ij}^{0}|\left(2L_{ij}\rho L_{ij}^{\dagger}-L_{ij}^{\dagger}L_{ij}\rho-\rho L_{ij}^{\dagger}L_{ij}\right).\label{eq:Mster_eq}
\end{align}

Here, we have identified $\xi_{0}|J_{ij}^{0}|=\frac{\zeta_{ij}^{2}}{\gamma \gamma_{c}}$. We also introduce
a time-dependent coefficient $\xi_{0}(\tau)$ to slowly vary the coupling
strength between the oscillators dynamically over time, where the
coherent coupling is given by the last term of the above equation.
The overall coupling is given by: 
\begin{equation}
J_{ij}\left(\tau\right)=\xi_{0}\left(\tau\right)J_{ij}^{0},
\end{equation}
where the $J_{ij}^{0}$ matrix describes the relative coupling strength
between the modes so that $J_{ij}^{0}=J_{ji}^{0}$, and we set $J_{ii}^{0}=J_{jj}^{0}=0$.
If the signs are the same throughout, the sign of $J_{ij}$ distinguishes
the ferromagnetic case $(+)$, in which the spins are aligned, from
the anti-ferromagnetic $(-)$ case.

In the classical limit, the corresponding mean-field (MF) equation
is \cite{inui2020entanglement}:
\begin{equation}
{
\frac{\partial\alpha_{i}}{\partial\tau}=-\alpha_{i}+\alpha_{i}^{*}\left(\lambda-g^{2}\alpha_{i}^{2}\right)+\sum_{k \neq i}{\left(J_{ki}\alpha_{k}-\left|J_{ki}\right|\alpha_{i}\right)}.\label{eq:Mean Field equation}}
\end{equation}
{In the mean-field simulations, this is initialized with a complex noise,
\begin{align}
\eta = \sigma_{\eta}\left(\eta_{1} + i\eta_{2}\right),
\label{eq:complex_noise}
\end{align}
where $\eta_{1}$ and $\eta_{2}$ are real gaussian noises with mean 0 and variance 1. The complex noise $\eta$ therefore has a total variance $<|\eta|^2> = 2\sigma_{\eta}^2$.}

Eq. \ref{eq:Mean Field equation} is known to provide a classical
route to finding the Ising ground state, apart from issues with solutions
being trapped in non-extremal states. For a coupled system, if we
assume that $\alpha_{i}=\sigma_{i}\left|\alpha\right|$, with the
same mean photon number for all modes at the steady state, then the
dynamics is such as to minimize $E\left(\bm{\sigma}\right)$, and
in the steady state
{
\begin{align}
\left|\alpha\right|^{2} & \sim\frac{\lambda-1}{g^{2}}.
\end{align}
}
We next consider the full quantum problem, and investigate conditions
for finding the solution in regimes where the mean-field equations
are no longer applicable, due to strong entanglement and quantum noise
effects. We then  compare this with the classical MF results.

\subsection{Success probability measurement}

We wish to evaluate the success probability, which is the probability
of measuring the system in an optimal solution. Ideally, one should
compute the external measurement at the quadrature detector, using
input-output theory \cite{Collett_PRA1984,drummond2004quantum,gardiner2004quantum,he2009dynamical,teh2017simulation,Teh2018},
since this is what is actually measured. In this paper we calculate
the quadrature probabilities of the internal field operators. We have
verified that in the final steady state these closely match the externally
measured success rates, which will be explained in a subsequent publication \cite{Teh_homodyne_2026}.

The goal of calculating the internal operator success rate requires
calculating the joint probabilities $P(x_{1},x_{2},...x_{M})$ of
internal quadratures, where $x_{i}\ge0$ corresponds to spin-up and
$x_{i}<0$ corresponds to spin-down for the $i^{th}$ mode. For simplicity,
we first consider a system described by the single-mode density matrix
$\rho=|\psi\rangle\langle\psi|$. The state $|\psi\rangle$ is expanded
in the Fock (number) basis as $|\psi\rangle=\underset{n}{\sum}\psi_{n}|n\rangle$,
where $\psi_{n}$ are the expansion coefficients. Hence, the density
matrix can be written as
\begin{align}
\rho & =\underset{n,m\ge0}{\sum}\rho_{nm}|n\rangle\langle m|,
\end{align}
 where $m,n=0,1,...$. We expand the density matrix in the $n$-photon
basis $|n\rangle$, with coefficients $\rho_{nm}$.
The $x$-quadrature probability distribution can be computed from
the density matrix $\rho$, expanded in the number state basis for small enough photon numbers:
\begin{align}
\langle x|\rho|x\rangle & =\underset{n,m=0}{\sum^{N_c}}\rho_{nm}\langle x|n\rangle\langle m|x\rangle,
\end{align}
where $|x\rangle$ is the $x-$quadrature eigenstate, and the photon number cutoff is $N_c$. From the diagonal elements of $\langle x|\rho|x\rangle$
, we can evaluate the probability density $P(x)$ of the $x$-quadrature
measurement .

Next, we calculate the joint probability distribution for the quadrature
measurement of a DOPO with $M$ modes:
\begin{align}
P(x_{1},x_{2},...x_{M}) & =|\langle\psi|x_{1},x_{2}...x_{M}\rangle|^{2}\nonumber \\
 & =\underset{\bm{m},\bm{n}=0}{\sum^{N_c}}\psi_{\bm{m},\bm{n}}\underset{i=1}{\overset{M}{\prod}}\langle m_i|x_{i}\rangle\langle x_{i}|n_i\rangle.\label{eq:joint_quad_prob}
\end{align}

Here $\bm{m}\equiv\left[m_{1},m_{2},..,m_{M}\right]$ and $\bm{n}\equiv\left[n_{1},n_{2},..,n_{M}\right]$.
To evaluate the success probability $P_{\bm{\sigma}}$ of observing
a spin vector $\bm{\sigma}$ from the joint probabilities $P(x_{1},x_{2},...x_{M})$,
we integrate over the quadratures:
\begin{align}
P_{\bm{\sigma}}= & \int_{\sigma_{1}}\int_{\sigma_{2}}...\int_{\sigma_{M}}P(x_{1},x_{2},...x_{M})dx_{1}dx_{2}...dx_{M}\,.
\end{align}
The integration domains are defined so that $\int_{+1}$ is an integral
from $0$ to $\infty$, and $\int_{-1}$ is an integral from $-\infty$
to $0$. This can be written in the form:
\begin{align}
P_{\bm{\sigma}}= & \int_{\bm{D}}P(\bm{x})d^{n}\bm{x}\nonumber \\
 & =\underset{\bm{m},\bm{n}}{\sum}\psi_{\bm{m}\bm{n}}\underset{i=1}{\overset{M}{\prod}}\int_{\sigma_{i}}\langle m|x_{i}\rangle\langle x_{i}|n\rangle dx_{i}\nonumber \\
 & =\underset{\bm{m},\bm{n}}{\sum}\psi_{\bm{m}\bm{n}}\underset{i=1}{\overset{M}{\prod}}\varLambda_{m_{i}n_{i}}^{\sigma_{i}}.
\end{align}

On binning the quadrature measurements for two possible spins, we
have $\sigma_{i}=\{1,-1\}$ depending on the spin of each mode. When
measuring the success probability of our simulations, we utilize the
tensor identity $\left(A\otimes B\otimes...\right)=\left(A\otimes I\otimes...\right)\left(I\otimes B\otimes...\right)...$,
allowing the algorithm to scale efficiently with the wave-function.

In this expression, $\varLambda_{m_{i}n_{i}}^{\sigma_{i}}$ are the
integrals for the $i^{th}$ mode, for a detected spin of $\sigma_{i}=\pm1$,
evaluated analytically:
\begin{align}
\varLambda_{mn}^{\sigma} & =\int_{\sigma}H_{m}(x)H_{n}(x)e^{-x^{2}}dx\nonumber \\
 & =\frac{m!n!}{2}\underset{p=0}{\overset{\left\lfloor \frac{m}{2}\right\rfloor }{\sum}}\underset{l=0}{\overset{\left\lfloor \frac{n}{2}\right\rfloor }{\sum}}\frac{\left(2\sigma\right){}^{2\left(\bar{m}-p-l\right)}\Gamma\left(\bar{m}-p-l+\frac{1}{2}\right)}{p!l!(m-2p)!(n-2l)!(-1)^{p+l}}.
\end{align}
Here $\bar{m}\equiv(m+n)/2$, and $\left\lfloor \frac{m}{2}\right\rfloor $,
$\left\lfloor \frac{n}{2}\right\rfloor $ indicate the floor function
which rounds down its argument to the nearest integer. We use the
standard gaussian integral result:
\begin{align}
\int_{\sigma}x^{n}e^{-x^{2}}dx & =\sigma^{n}\frac{\Gamma(\frac{n+1}{2})}{2}.
\end{align}

\subsection{Purity Calculation \label{subsec:Purity-Calculation}}

We can calculate the quantum purity in our simulation as a means of
determining the effects of decoherence of the initial quantum state.
The sampled purity of the conditional density matrix $\rho$ can be
evaluated from the sampled trajectory wave-functions $|\psi_{i}\rangle$, as explained in Appendix A,
given by:
\begin{align}
\rho & =\frac{1}{N_{s}}\overset{N_{s}}{\underset{i=1}{\sum}}|\psi_{i}\rangle\langle\psi_{i}|,\label{eq:conditional_rho}
\end{align}
where $N_{s}$ is the number of sampled trajectories. The squared density
matrix $\rho^{2}$ is:
\begin{align}
\rho^{2} & =\frac{1}{N_{s}^{2}}\overset{N_{s}}{\underset{i,j=1}{\sum}}|\psi_{i}\rangle\langle\psi_{i}|\psi_{j}\rangle\langle\psi_{j}|,
\end{align}
Defining $|\psi_{i}\rangle\langle\psi_{i}|=\underset{\bm{n},\bm{m}}{\sum}\rho_{\bm{nm}}^{(i)}|\bm{n}\rangle\langle \bm{m}|$,
the purity is given by the trace of $\rho^{2}$ , expressed as:
\begin{align}
Tr(\rho^{2}) & =\frac{1}{N_{t}^{2}}\overset{N_{t}}{\underset{i,j=1}{\sum}}\underset{\bm{n},\bm{m}=0}{\sum^{N_{c}}}\rho_{\bm{nm}}^{(i)}\rho_{\bm{mn}}^{(j)}.
\end{align}

Here $N_{c}$ is the cut off, $M$ is the number of modes and $\rho_{\bm{nm}}^{(i)}$
are the $i$-th density matrix elements. In evaluating the summation
in $Tr(\rho^{2})$, the indices $\bm{n}$ and $\bm{m}$ are vectors, with each component running over $0,1,2,...N_{c}$.

\section{Fixed parameter simulations}

In this section, we treat numerical simulations with fixed parameters.
A later section investigates dynamical, time-varying parameters which
can vary throughout the passage to the final, measured state.

\subsection{Initial quantum states \label{subsec:Initial-quantum-states}}

The initial state of our simulation should ideally encompass all possible
solutions, meaning that at the beginning, for a system with $M$ modes
and a particular spin configuration, the success probability would
be $\sim\frac{1}{2^{M}}$. The system is simulated in a truncated
Fock basis up to $N_{c}$, where $N_{c}>n_{c}$, where $n_{c}$ is
the average photon number, which largely determines how many modes
can be treated. For good accuracy, this must be large enough so that
errors due to number truncation are negligible. Yet, if the average
photon number $n_{c}$ is too low, the quantum noise causes prohibitive
errors in detecting the output quadrature sign.

A vacuum state in all modes $\psi_{vac}=|0\rangle^{\otimes M}$ is
the simplest to generate experimentally. This can evolve to any solution
at later times. We therefore initialize some quantum simulations
with a vacuum state and use these as a benchmark. One of the ways
we investigate the influence of quantum effects is by initializing
the simulation with more non-classical quantum superposition states.
The motivation is to explore whether such a state can influence the
simulation's performance. In particular, we treat highly entangled
initial cat states, under the assumption that these extremely non-classical
states may provide evidence of any quantum advantage.

We consider two alternative initial quantum states: the first is an
outer product of cat states equivalent to all $2^{M}$ spin states,
\begin{align}
|\psi_{sup}\rangle & =\mathcal{N}_{c}^{M}\left(|\alpha\rangle+|-\alpha\rangle\right)^{\otimes M}\label{eq:psi_sup}
\end{align}
with coherent amplitude $\alpha$. The second state is an $M$-partite
entangled state:
\begin{align}
|\psi_{ent}\rangle & =\mathcal{N}_{M}\left\{ \left(|\alpha\rangle+|-\alpha\rangle\right)_{1}|0\rangle_{2}..|.0\rangle_{M}\right.\nonumber \\
 & +|0\rangle_{1}\left(|\alpha\rangle+|-\alpha\rangle\right)_{2}..|.0\rangle_{M}+...\nonumber \\
 & \left.+|0\rangle_{1}|0\rangle_{2}...\left(|\alpha\rangle+|-\alpha\rangle\right)_{M}\right\} \label{eq:psi_ent}
\end{align}
where $\mathcal{N}_{M}$ is the normalization factor. This is a quantum
superposition, but with only one mode initialized to a cat state.
The rationale for using $\psi_{sup}$ and $\psi_{ent}$, is to investigate
whether introducing quantum superposition and entanglement at the
start of the simulation improves the CIM performance compared to initializing
with a vacuum state $\psi_{vac}$. 

{The quantum states in Eqs. \ref{eq:psi_sup}
and \ref{eq:psi_ent} contain the Schrodinger cat state in each mode. The cat state can be generated in a DOPO if the nonlinear gain saturation parameter $g$ is larger than one. To the best of our knowledge, the largest optical nonlinear gain value is found in the InGaP quantum nanophotonic integrated circuits \cite{Zhao:22}. However, this gain is two orders of magnitude less than unity. On the other hand, $g>1$ has been achieved in the superconducting circuit experiments \cite{Leghtas_Science2015,Reglade:2024aa,Beaulieu:2025aa}.}

\begin{figure}
\centering{}\includegraphics[width=0.6\columnwidth]{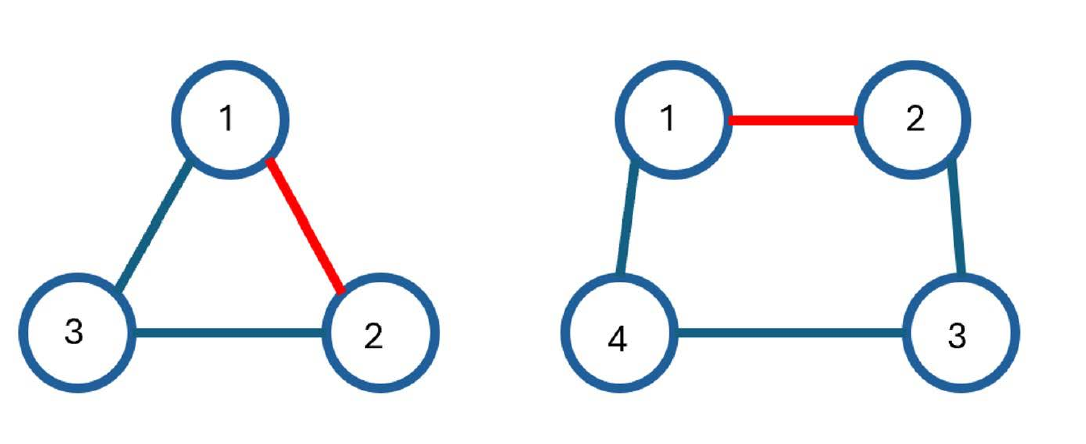}\caption{Ising spin diagram of $M=3$ and $M=4$ spin arrangements
with nearest neighbor interactions only, with uniform interaction
strength ($J_{ij}=-1$, where $i$ and $j$ are nearest neighbor modes),
with the sign of the $J_{12}$ interaction flipped ($J_{12}=J_{21}=1$).
{.} \label{fig:Diag}}
\end{figure}

\begin{figure}
\begin{centering}
~\includegraphics[width=0.5\columnwidth]{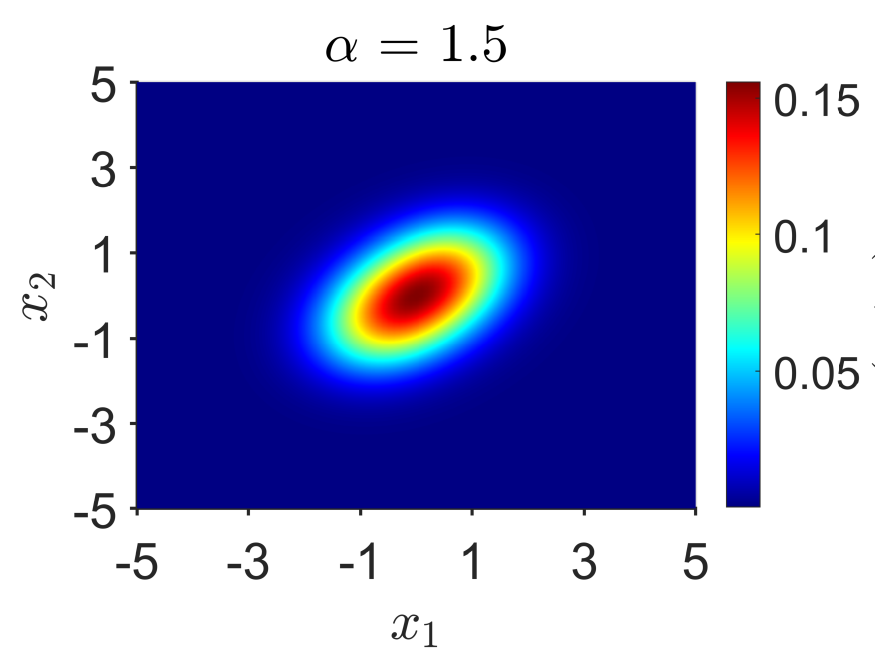}
\par\end{centering}
\begin{centering}
\includegraphics[width=0.5\columnwidth]{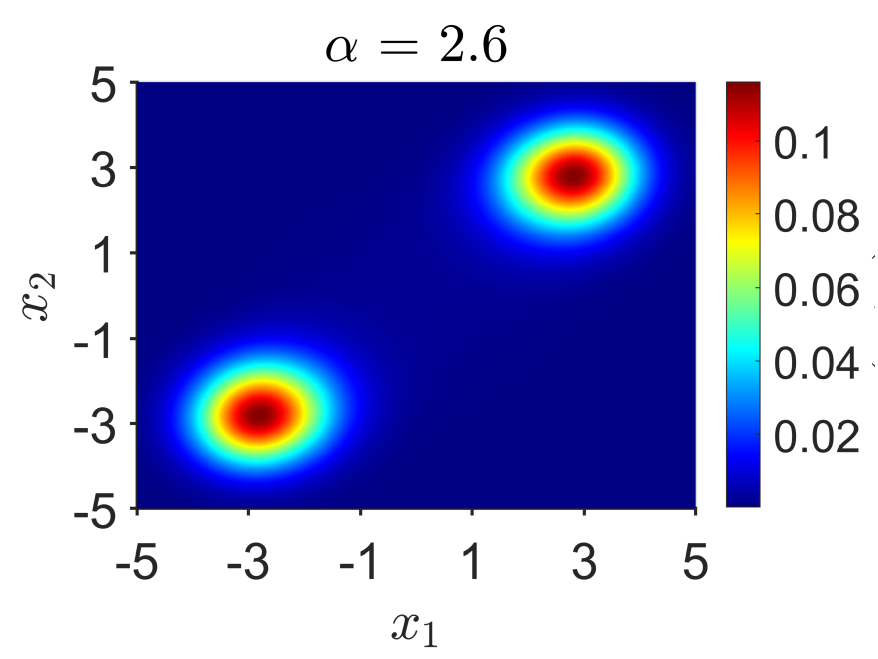}
\par\end{centering}
\caption{Snapshot of the joint probability distribution for the quadrature
measurement $P(x_{1},x_{2})$, as defined in Eq. \ref{eq:joint_quad_prob},
shown for two cases top: $\alpha=1.5$ and bottom: $\alpha=2.6$,
in a two-mode coupled DOPO system at steady state, $\tau=4$. The
simulation parameters are set as $\xi_{0}=0.5$, $\lambda=\alpha^{2}g^{2}$,
with $g=0.6$ and time-independent coupling matrix elements $J_{ij}=1$
when $i\protect\neq j$, and $J_{ii}=0$. The system was
initialized in the vacuum state $\psi_{vac}$, with a  photon number cutoff of $N_{c}=16$. \label{fig:Surf_alpha}}
\end{figure}

\begin{figure}
\begin{centering}
~\includegraphics[width=0.5\columnwidth]{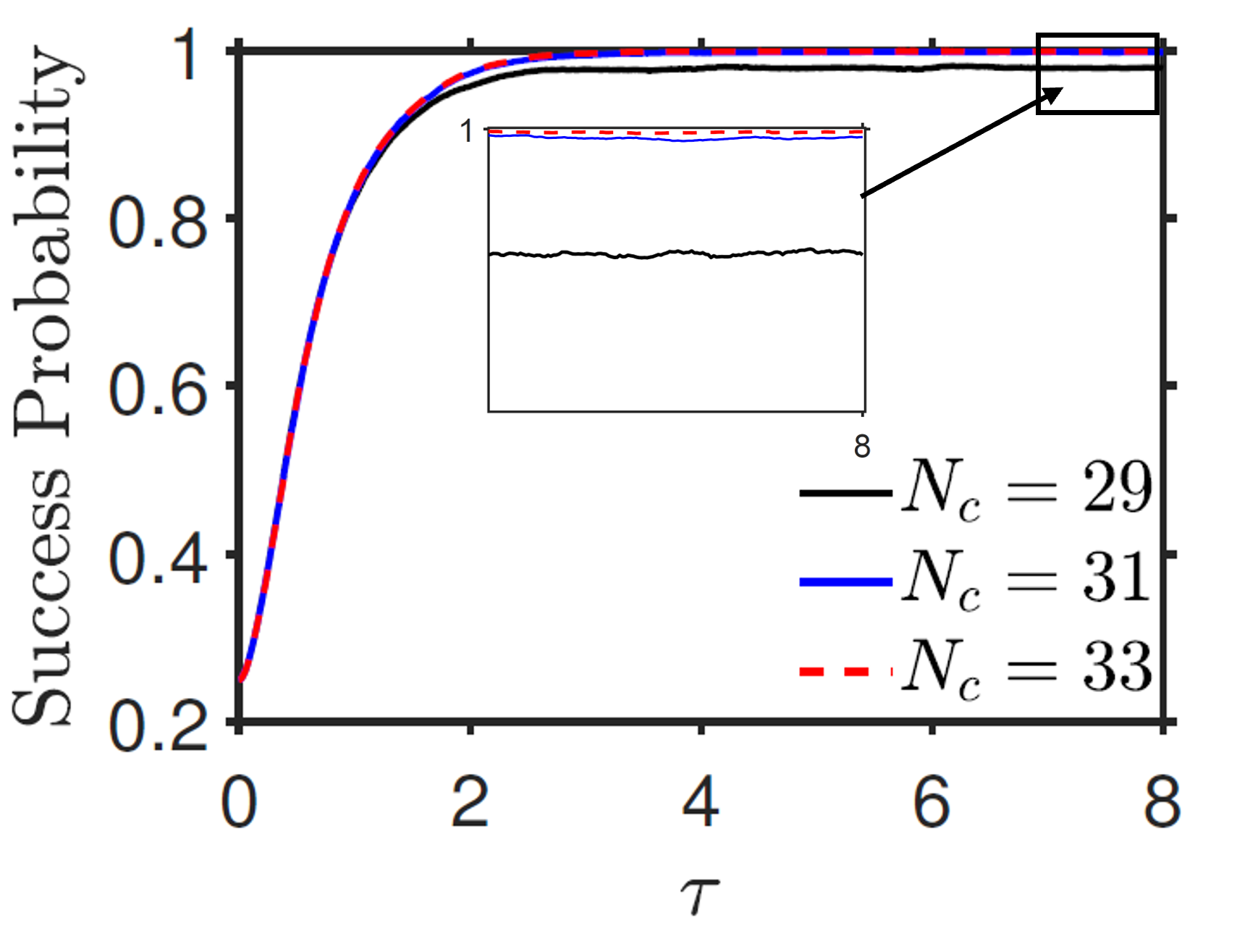}
\par\end{centering}
\caption{Success probability variation with varying photon cut-off $N_{c}$.
The simulation parameters are $\xi_{0}=0.5$, $\lambda=5.4$ and
$g=0.6$. Here the couplings $J_{ij}$ are as in the $M=3$ case in
Fig (\ref{fig:Diag}), with $\psi_{vac}$ as the initial state. {.}
\label{fig:Success_vs_N}}
\end{figure}

\subsection{Antiferromagnetic impurity problems\label{subsec:Case-I:-Frustrated}}

We first consider a simple case of antiferromagnetic Ising spins ($J_{ij}<0$),
where neighboring anti-parallel spins result in the lowest energy
with uniform interaction strength $|J_{ij}|$ (the same for all $i\neq j$)
and $J_{ij}=0$ when $i=j$. We examine two cases as in Fig. \ref{fig:Diag},
each with an impurity which flips the sign of the interaction. The
first involves spins arranged in a triangle ($M=3$) and the second involves
spins arranged in a square ($M=4$).

In the first case, the energy is minimized when each spin is aligned
opposite to its neighbors, apart from the impurity. We take $J_{ij}=-1$,
where $i$ and $j$ are nearest neighbor modes. After the first two
spins are aligned oppositely, the third spin becomes \textit{frustrated}
since there is no configuration that provides the same low energy.
This simultaneous minimization with respect to the other two spins
leads to a six-fold degeneracy in a uniform anti-ferromagnet. In this
study, we flip the sign of $J_{12}$ ($J_{12}=J_{21}=1$) interaction
as given in table \ref{tab:Fixed-parameters} , thereby reducing the
degeneracy to two and curing the frustration. We also include the case of $M=4$, where the frustration is \emph{caused}
by the impurity.

\begin{figure}
\begin{centering}
\includegraphics[width=0.4\columnwidth]{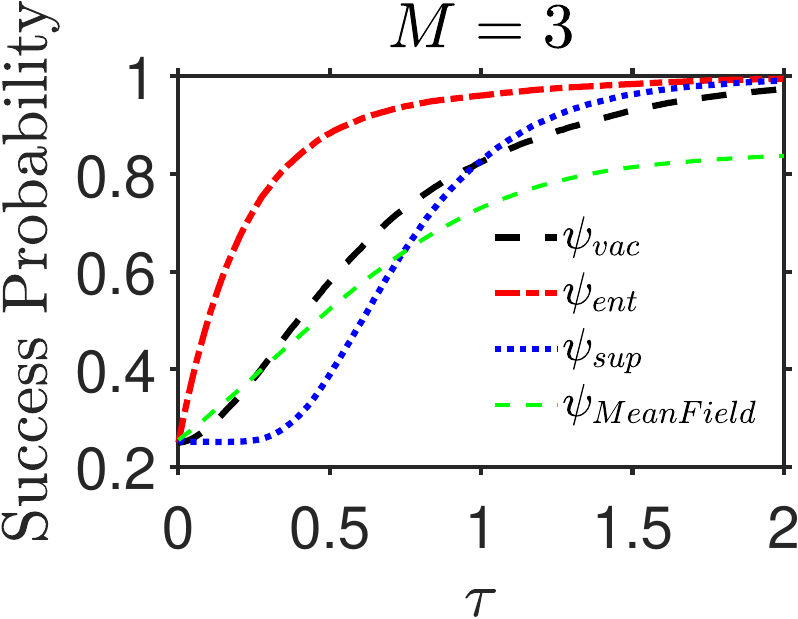}$\,$\includegraphics[width=0.4\columnwidth]{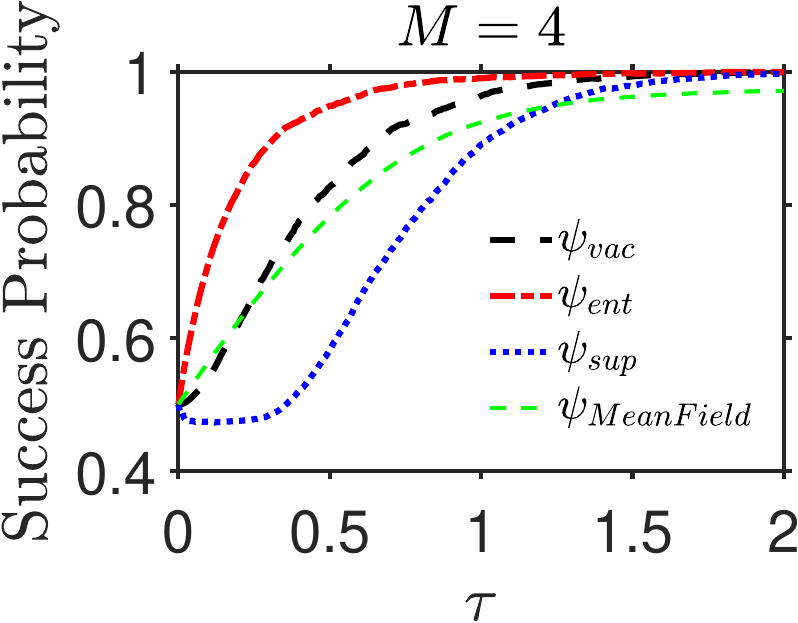}
\par\end{centering}
\caption{Evolution of the system dynamics for the frustrated anti-ferromagnetic
problem. Top: Success probability and bottom: total photon number
summed over all modes, shown for initial states with increasing coherent
amplitude to $\alpha = 3.87$, achieved by raising the pump strength to $\lambda=5.4$. A photon cut-off of $N_{c}=31$
is used. The rest of the simulation parameters are fixed at $\xi_{0}=0.5$
and $g=0.6$ . The couplings $J_{ij}$ correspond to the $M=3$ and
$M=4$ configurations shown in Fig. (\ref{fig:Diag}), with simulation
time $\tau=2$ and $\tau_{steps}=300$, averaged over $10^{5}$ realizations.
The quantum superposition results (red and blue) outperform both the
vacuum-initialized system labeled by $\psi_{vac}$ (black), and the
mean-field approximation $\psi_{Mean\,Field}$ (green) obtained from
{ Eq.\ref{eq:Mean Field equation} using an initial complex noise with $\sigma_{\eta} = 0.1$.} \label{fig:vary_alpha}}
\end{figure}

\begin{table*}
\begin{centering}
\begin{tabular}{|c|c|c|}
\hline 
$M=3$ & $M=4$\tabularnewline
\hline 
\hline 
\begin{cellvarwidth}[t]
\centering
${\lambda=5.4}$\medskip{}

$g=0.6$\medskip{}

$\xi_{0}=0.5$\medskip{}

$\tau_{max}=2$\medskip{}

$J=\left[\begin{array}{ccc}
0 & 1 & -1\\
1 & 0 & -1\\
-1 & -1 & 0
\end{array}\right]$
\end{cellvarwidth} & \begin{cellvarwidth}[t]
\centering
${\lambda=5.4}$\medskip{}

$g=0.6$\medskip{}

$\xi_{0}=0.5$\medskip{}

$\tau_{max}=2$\medskip{}

$J=\left[\begin{array}{cccc}
0 & 1 & 0 & -1\\
1 & 0 & -1 & 0\\
0 & -1 & 0 & -1\\
-1 & 0 & -1 & 0
\end{array}\right]$
\end{cellvarwidth}\tabularnewline
\hline 
\end{tabular}
\par\end{centering}
\caption{Parameters for the fixed parameter simulations 
in Fig. \ref{fig:vary_alpha}. \label{tab:Fixed-parameters}}
\end{table*}

We simulate each system starting from an initial vacuum state $\psi_{vac}$
and, for comparison purposes, benchmark it against the initial states
$\psi_{sup}$ and $\psi_{ent}$, as defined by Eqs. \ref{eq:psi_sup}
and \ref{eq:psi_ent}. For large values of the coherent amplitude
$\alpha=\sqrt{\lambda}/g$, the quadrature components become well
separated in quadrature distribution space, at steady state. Specifically,
for $\alpha>2.5$ we observe clear quadrature separation, as illustrated
in Fig. \ref{fig:Surf_alpha} for the simple case of two coupled DOPOs.
Any overlap of quadrature solutions, as seen by Top plot in Fig. \ref{fig:Surf_alpha},
could lead to a reduction of success probability at the ground state,
hence an under performance.

Due to computational limitations in CPU and memory, we fix $\alpha=3.8730$
throughout these simulations, which corresponds to a steady-state
mean photon number of approximately $15$. We choose a photon number
cutoff of $N_{c}=31$, including the vacuum state $|0\rangle$, which
is sufficient as it spans three standard deviations from the mean
photon number. The adequacy of this cutoff is validated in Fig. \ref{fig:Success_vs_N},
where we vary the photon cutoff $N_c$ to assess the associated information
loss, represented by the success probability.

As shown in Fig. \ref{fig:vary_alpha}, the cases with $M=3$ demonstrate
a clear performance advantage when the simulation is initialized with
a quantum state such as $\psi_{sup}$ and $\psi_{ent}$, compared
to a vacuum state $\psi_{vac}$, as they reach the peak success probability
more rapidly. The estimated time-step and sampling errors for success probability in Fig.
\ref{fig:vary_alpha}, remain below $0.3\%$. The expressions used to
estimate the time-step and sampling errors are provided in the appendix.
However, as the number of modes increases, this performance advantage
diminishes between the quantum simulations. To understand this behavior, we examine the master equation,
Eq. \ref{eq:Mster_eq}, where the last three terms---including the
coupling terms---are dissipative in nature. Introducing $M$ modes
adds $3M$ dissipative terms to the master equation in systems with
only nearest-neighbor coupling. When all modes are mutually coupled---including
self-coupling---the number of dissipative terms become $M(M+2)$.
Consequently, as $M$ increases, the cumulative dissipation grows,
rapidly suppressing quantum effects at the onset of the simulation.
As a further investigation, we conduct a purity analysis to quantify
how quickly the quantum state undergoes decoherence. This is studied
in detail in Sec. \ref{subsec:Purity-Analysis}.

From Fig. \ref{fig:vary_alpha} , we observe an improvement in performance
for all cases ($M=3$ and $M=4$) , when a non-classical state
is used at the start of the simulation. Although the coherent superposition
state $\psi_{sup}$ performs poorly initially, it eventually surpasses
the initial vacuum state $\psi_{vac}$ in the $M=3$ case. This suggests that
both entanglement and superposition may contribute to enhancing the
performance of the coherent Ising machine (CIM). 

\subsection{Mean-field approximation vs full quantum dynamics}

Next, we compare the classical mean-field predictions obtained from
Eq. \ref{eq:Mean Field equation} with the full quantum simulations.

In our first example, we consider the impure antiferromagnetic Ising problem introduced in Fig. \ref{fig:Diag}, together with the corresponding
quantum results shown in Fig. \ref{fig:vary_alpha}. To ensure a balance between physical accuracy and available computational resources, we restrict the photon number to a feasible cutoff. Increasing the coherent amplitude $\alpha$ of the initial quantum states necessitates a corresponding increase in the pump strength $\lambda$ thereby approaching more realistic experimental conditions, where larger coherent amplitudes and higher photon numbers are typically encountered. Under these conditions, accurately capturing the full quantum dynamics requires a photon cutoff significantly larger than $N_{c}=31$ for each mode, which would substantially increase the computational cost of the quantum simulations.

We compare two quantum initializations, corresponding to Eqs. \ref{eq:psi_sup}
and \ref{eq:psi_ent}, denoted $\psi_{sup}$ and $\psi_{ent}$, respectively.
The entanglement in $\psi_{ent}$ at $\tau=0$ is verified using the
positive partial transpose (PPT) criterion \cite{seperability},
which provides a necessary condition for entanglement when negative
eigenvalues $\lambda_{i}$ appear in the partially transposed density
matrix. We compute the resulting negativity $\mathcal{\mathbb{N}}=\underset{\lambda_{i}<0}{\sum}|\lambda_{i}|$
and obtain $\mathcal{\mathbb{N}}\sim0.47$, for sufficiently large
$\alpha>0$, confirming that $\psi_{ent}$ is indeed entangled.

The results show a clear performance enhancement, the success probability
(top panel of Fig. \ref{fig:vary_alpha}), approaches $>95\%$ for
$t<1$, for the large $\lambda$ case. In the $M=4$ case, in addition
to a speed up, we observe a significant improvement of $\sim5\%$,
in the success probability, compared to the classical counterparts.

The results are compared against two benchmark systems to highlight
the quantum advantage obtained when the system is prepared in the
entangled state $\psi_{ent}$, shown by the red curves in Fig. \ref{fig:vary_alpha}.
The first benchmark is the quantum simulation initialized with a vacuum
state, $\psi_{vac}=|0\rangle^{\otimes M}$, simulated with pump strength
$\lambda=5.4$ (solid black) in Fig. \ref{fig:vary_alpha}.

{ In the second benchmark, we compare with results labeled $\psi_{Mean\,Field}$,
obtained from the mean-field approximation (Eq. \ref{eq:Mean Field equation}).
In Fig. \ref{fig:vary_alpha}, we choose $\sigma_{\eta} = 0.1$ which gives a complex noise variance of $0.02$, using parameters $\xi_{0}=0.5$, $g=0.6$ and $\lambda=5.4$.}
This gives a slower approach to the solution, and a lower success
probability. Together, these comparisons suggest that if experiments
can drive the CIM into a truly quantum regime with large photon numbers,
significant performance gains can be realized over the classical systems.

\subsection{Quantum tunneling out of false mimima}

We next show that in cases where the classical trajectory becomes trapped
in a false minimum, the success rate is greatly increased in the quantum
regime compared to the mean field predictions. This is a strong indication of potential quantum advantage, since trapping in false minima is a known classical problem

\begin{figure}
\begin{centering}
\includegraphics[width=0.4\columnwidth]{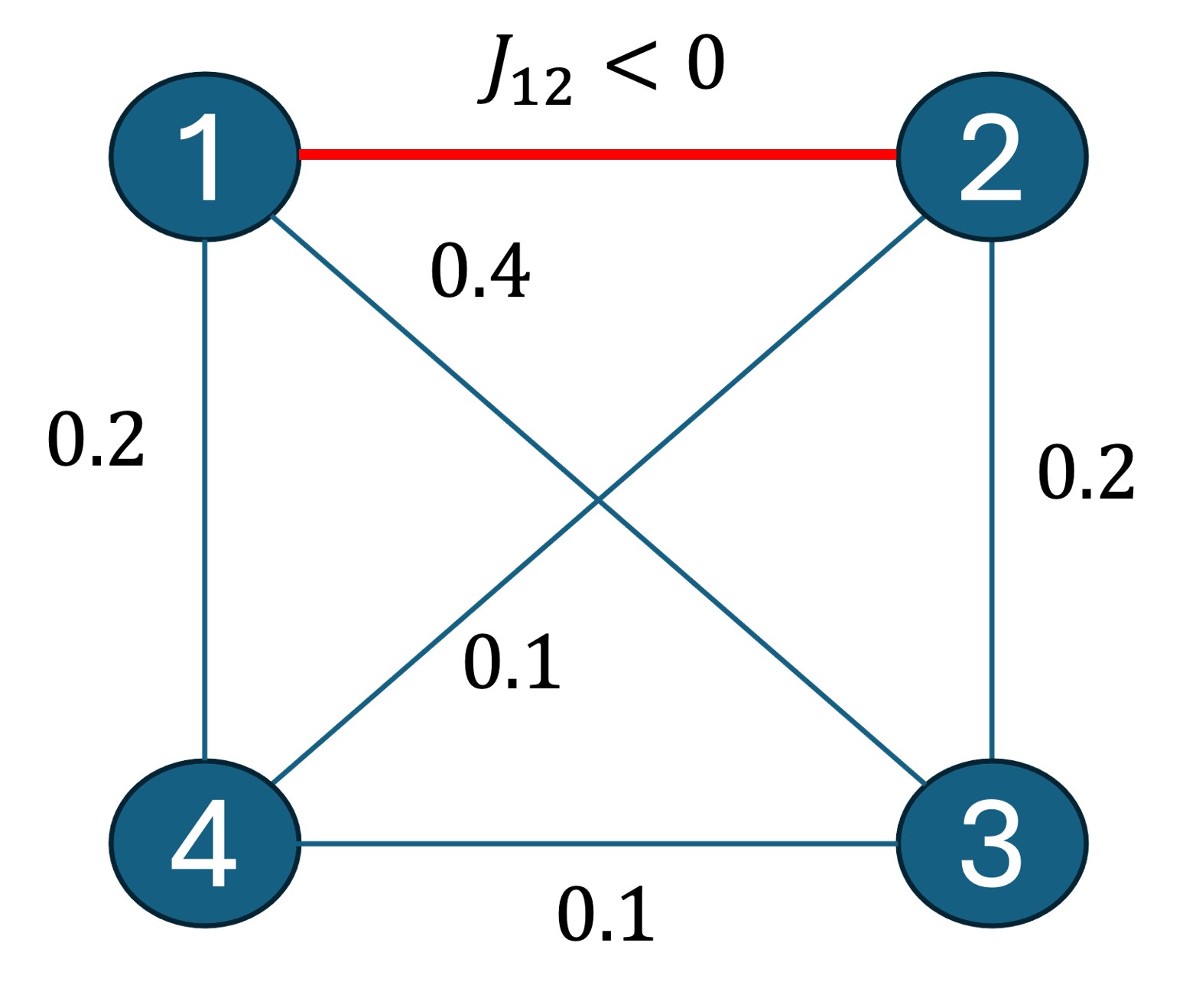}
\par\end{centering}
\caption{Ising spin diagram for the ferromagnetic spin problem with specialized
weights. \label{fig:Diag-1}}
\end{figure}

\begin{figure}
\begin{centering}
\includegraphics[width=0.5\columnwidth]{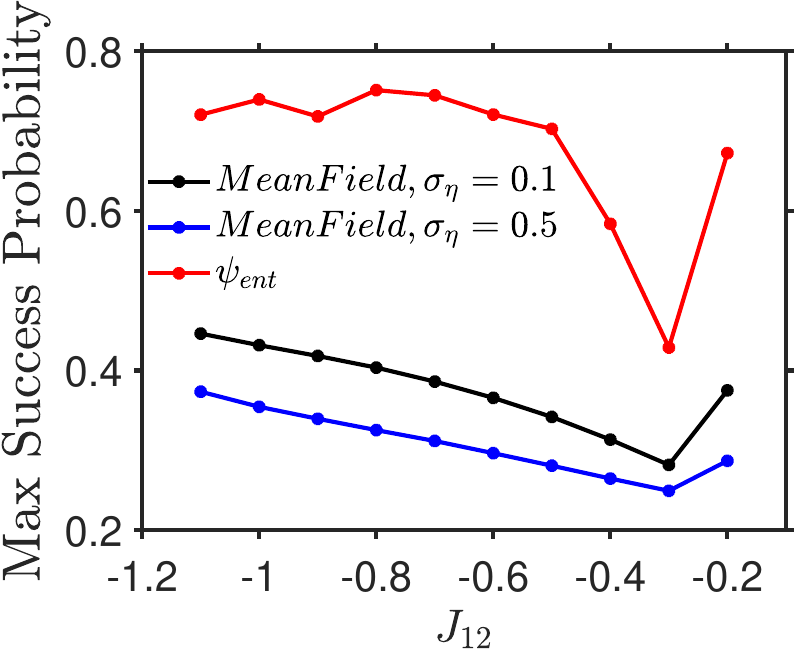}
\par\end{centering}
\caption{{Maximum success probability as a function of $J_{12}$ for the weighted ferromagnetic Ising configuration shown in Fig.~\ref{fig:Diag-1}, using the initial quantum state $\psi_{ent}$. The results are compared with those obtained from the mean-field approximation. Simulation parameters are $\lambda=2.4$, $g=0.4$, $\xi_{0}=0.5$, and coherent amplitude $\alpha = 3.87$ for the initial quanntum state $\psi_{ent}$. The simulations were performed up to $\tau = 16$ with $\tau_{steps}=2400$, and a photon cutoff of $N_{c} = 31$. The mean-field simulations were initialized using the complex noise $\eta$ defined in Eq. \ref{eq:complex_noise}. \label{fig:Mean_Field}}}
\end{figure}

In this example, we consider a weighted ferromagnetic Ising problem
illustrated in Fig. \ref{fig:Diag-1}, where we flip the sign of $J_{12}$.
This case is known to have coupling regimes where classical CIM solvers
become trapped in false minima \cite{Rah2023private}. In Fig. \ref{fig:Mean_Field},
we plot the maximum success probability achieved within the simulation
window $\tau_{max}=16$ while varying $J_{12}$. The mean-field model
exhibits a drop in performance around $J_{12}\approx-0.3$,
as shown by the black curve of Fig. \ref{fig:Mean_Field}. The observed
reduction in success probability around $J_{12}\approx-0.3$ in Fig.
\ref{fig:Mean_Field}, arises from an energy landscape containing
multiple near-optimal local minima, where the mean-field trajectories
become trapped.

The quantum simulations (red curve in Fig. \ref{fig:Mean_Field}) also exhibit a drop in performance around $J_{12}\approx-0.3$. However, they maintain better performance than their classical counterparts, even in this challenging region.
This improved behaviour suggests that the quantum dynamics exploits non-equilibrium tunnelling processes \cite{Drummond_PRA1989,Sun_NJP2019},
helping the system to escape metastable configurations and reach the
true optimal solution. This discrepancy highlights the limitations
of the mean-field approximation, specifically its inability to capture
essential quantum effects and demonstrates the potential for a quantum
advantage when comparing a strongly interacting quantum CIM with a
more classical one.

\section{Dynamical strategies\label{sec:dyn_str}}

In this section, we explore a time-dependent strategies aimed at
enhancing quantum performance further, by carefully manipulating coupling strength, starting from a initial vacuum state $\psi_{vac}=|0\rangle^{\otimes M}$,
which aligns more closely with current experimental capabilities.

\subsection{Time dependent coupling \label{subsec:Time-dependent-coupling}}

Here we consider a weighted ferromagnetic problem introduced in Fig.
\ref{fig:Diag-1}.  We use different simulation parameters: $g=0.4$,
$\lambda=2.4$ and we choose $\xi_{0}(\tau)$, as given in table \ref{tab:Time-Varying-coupling-parameters},
such that , yielding a maximum steady-state amplitude of $|\alpha|\sim3.873$.
This amplitude necessitates a photon number cutoff of $N_{c}=31$.

\begin{table}
\begin{centering}
\begin{tabular}{|c|}
\hline 
Time Varying Coupling Strategy\tabularnewline
\hline 
\hline 
\begin{cellvarwidth}[t]
\centering
$|\psi_{initial}\rangle=|0\rangle^{\otimes M}$\medskip{}

$\lambda=2.4$\medskip{}

$g=0.4$

\medskip{}

$\xi_{0}(\tau)=\begin{cases}
0.5\\
2\\
\frac{1}{2}(3\tau/\tau_{max}+1)\\
\frac{1}{2}(3\tanh(\tau)+1)\\
\frac{1}{2}(\frac{6}{1+\exp(-0.5(t-16))}+1)
\end{cases}$\medskip{}

$\tau_{max}=16$\medskip{}

$J=\left[\begin{array}{cccc}
0 & -0.9 & 0.4 & 0.2\\
-0.9 & 0 & 0.2 & 0.1\\
0.4 & 0.2 & 0 & 0.1\\
0.2 & 0.1 & 0.1 & 0
\end{array}\right]$
\end{cellvarwidth}\tabularnewline
\hline 
\end{tabular}
\par\end{centering}
\caption{Parameters for the time dependent coupling strategy 
in Fig. \ref{fig:a:-Ferro_vary_J}. \label{tab:Time-Varying-coupling-parameters}}
\end{table}

\begin{figure}[t]
\centering
\begin{minipage}{0.48\columnwidth}
    \centering
    \includegraphics[width=\linewidth]{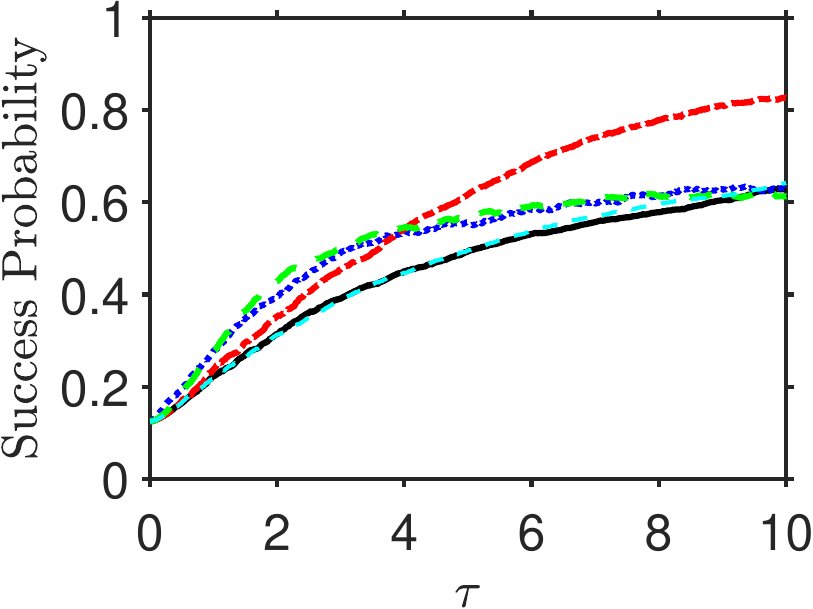}
\end{minipage}
\hfill
\begin{minipage}{0.48\columnwidth}
    \centering
    \includegraphics[width=\linewidth]{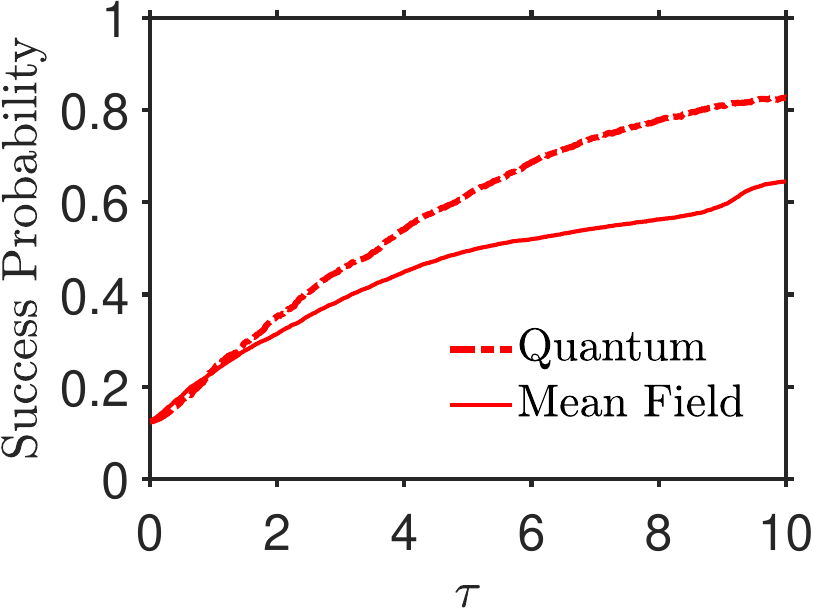}
\end{minipage}
\vspace{0.3cm}
\begin{minipage}{0.48\columnwidth}
    \centering
    \includegraphics[width=\linewidth]{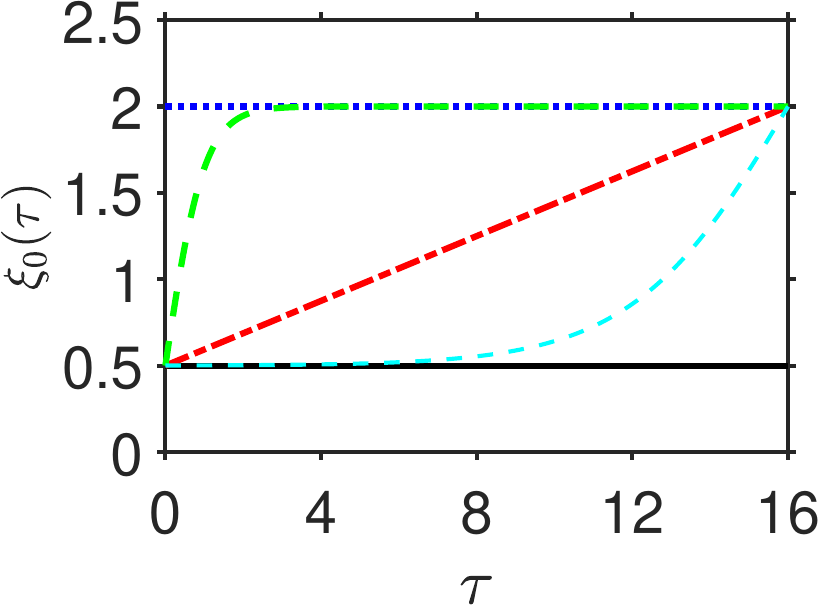}
\end{minipage}
\hfill
\begin{minipage}{0.48\columnwidth}
    \centering
    \includegraphics[width=\linewidth]{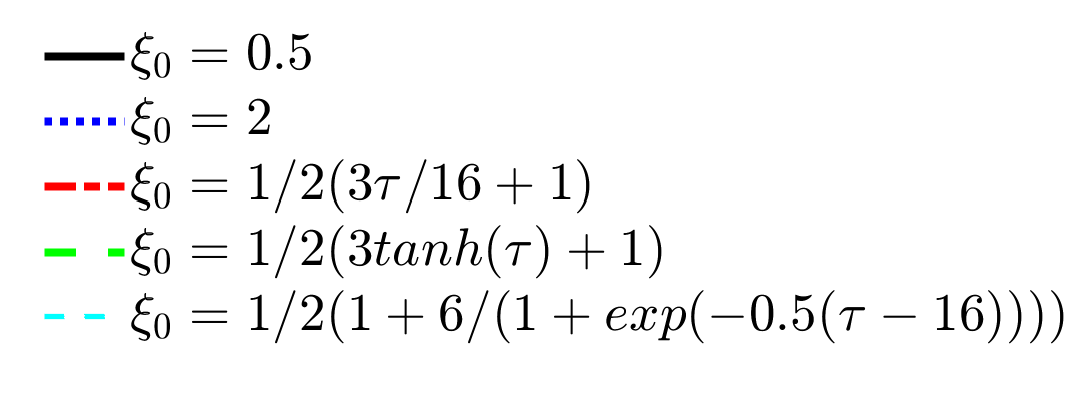}
\end{minipage}
\caption{{Top left: Time evolution of the success probability for the weighted ferromagnetic
problem as in Fig. \ref{fig:Diag-1} with $J_{12}= -0.9$,  using different
strategies for the interaction coefficient $\xi_{0}$, with initial
vacuum state $\psi_{vac}$. A photon cut-off of $N_{c}=31$ is used.
Top right: Quantum results compared against mean field results for the linear ramp case obtained from the mean-field equation in Eq. \ref{eq:Mean Field equation} using an initial complex noise with $\sigma_{\eta} = 0.1$.
Bottom: Comparison of four strategies for varying $\xi_{0}$ from
$\xi_{0}=0.5$ to $\xi_{0}=2$: constant (blue and black), a linear
increase (red), a hyperbolic tangent function to emulate a rapid
initial rise (green), and a slow incrrease initially followed by a rapid increase at a later time (cyan).}\label{fig:a:-Ferro_vary_J}}
\end{figure}

In Fig. \ref{fig:a:-Ferro_vary_J} $J_{12}$ is fixed at $-0.9$ and
simulations are performed using the vacuum initial state $|\psi_{vac}\rangle=|0\rangle^{\otimes M}$.
The interaction strength is varied dynamically by multiplying $J_{ij}$
with a time-dependent interaction coefficient $\xi_{0}(\tau)$ as
defined in Eq. \ref{eq:Mster_eq}.

The coefficient $\xi_{0}(\tau)$ is used to increase the interaction
strength by up to four times. Different strategies for increasing
$\xi_{0}(\tau)$ are shown in the bottom panel of Fig. \ref{fig:a:-Ferro_vary_J}.
Increasing $\xi_{0}(\tau)$ enhances the interaction between the DOPO
elements. Notably, when the simulation is initialized from the vacuum
state, the linear ramp strategy yields a substantial improvement in
the success probability---exceeding $30\%$ compared to other approaches
at simulation time $\tau=10$. This enhancement may not be unique
to the linear ramp itself; its effectiveness appears to be problem-dependent
and may not generalize across different configurations. The estimated
time-step and sampling errors associated with the results in top panel
of Fig. \ref{fig:a:-Ferro_vary_J} are below 2\%.

\subsection{Purity Analysis \label{subsec:Purity-Analysis}}

  { In this section, we investigate the loss of coherence in the highly quantum initial states $\psi_{sup}$ and $\psi_{ent}$, introduced in Sec. \ref{subsec:Initial-quantum-states} via Eqs. \ref{eq:psi_sup} and \ref{eq:psi_ent}, respectively. To quantify this coherence loss, we employ the purity measure defined in Sec. \ref{subsec:Purity-Calculation}. We consider two distinct nonlinearity regimes to enable a comparative analysis, allowing us to assess how non-linear loss ($g$) correlates with improved simulation performance.}

\begin{figure}[H]
\centering\includegraphics[width=0.35\columnwidth]{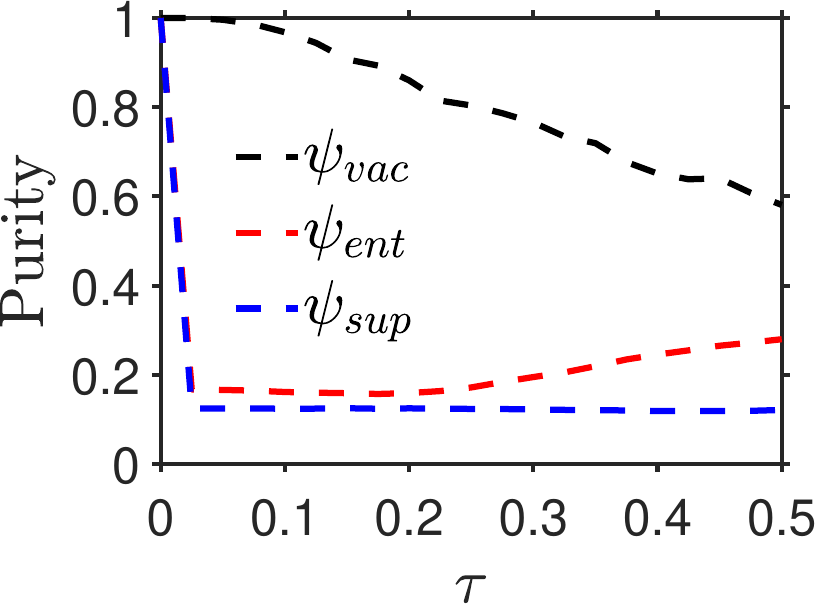}\includegraphics[width=0.35\columnwidth]{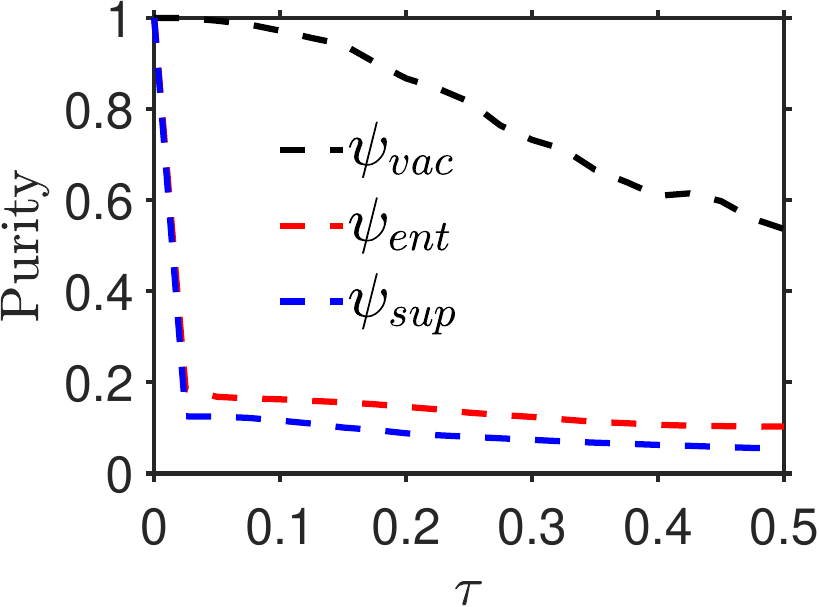}

\includegraphics[width=0.35\columnwidth]{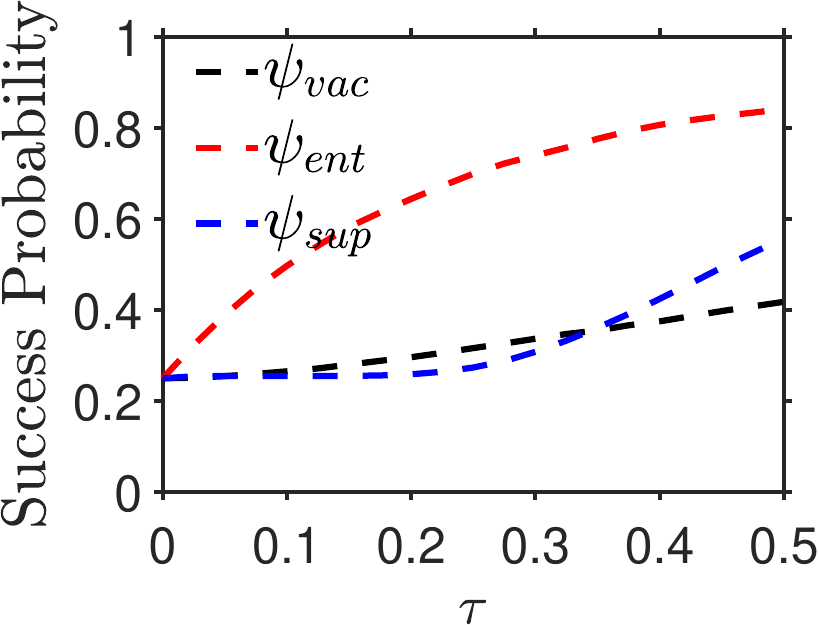}\includegraphics[width=0.35\columnwidth]{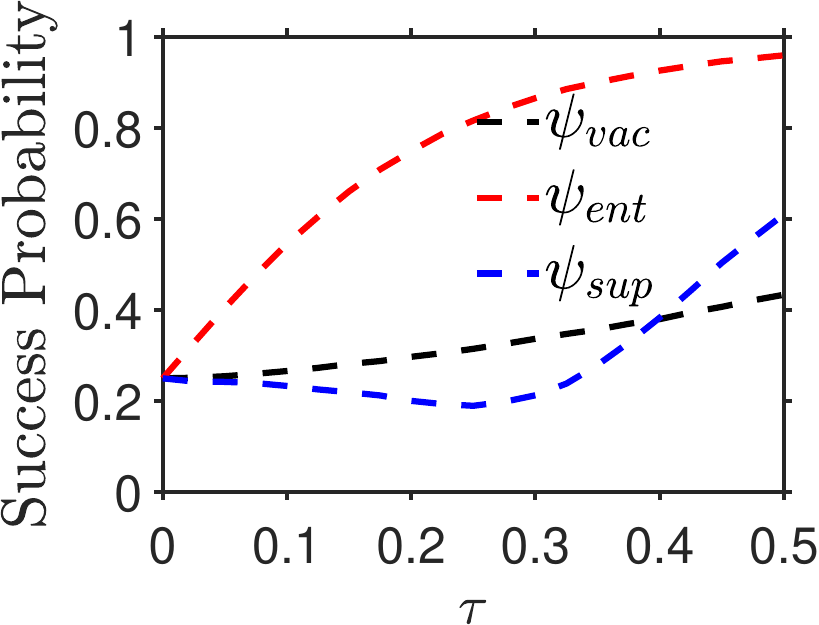}

\caption{Investigating the impact of initial states on purity and success probability
for $M=3$ case in Fig (\ref{fig:Diag}), with different initial states.
Top plots: purity and bottom plots: corresponding success probabilities.
Here we consider two settings, left plots: high nonlinearity setting
($g=0.6$) and right plots: low nonlinearity setting
($g=0.1$), where  $\xi_{0} = 0.5$, $\lambda = 2.4$, $\alpha = 3.87$  
with simulation times of $\tau_{max}=0.5$ with $\tau_{steps}=50$
using $10^{3}$ trajectories. Here a photon cut-off of $N_{c}=31$ is used. \label{fig:Purity}}
\end{figure}

 { The results for the three-mode impurity problem shown in Fig. \ref{fig:Diag} are obtained in a high-nonlinearity regime ($0 \ll g \leq 1$), where multiple damping channels contribute significantly to the master equation in Eq. \ref{eq:Mster_eq}. In this regime, coherence in highly quantum initial states decays rapidly. This behaviour is evident in the top panels of Fig. \ref{fig:Purity}, where the purity of $\psi_{ent}$ and $\psi_{sup}$ decreases at a faster rate compared to the vacuum state $\psi_{vac}$.}

\begin{figure}
\begin{centering}
\centering\includegraphics[width=0.4\columnwidth]{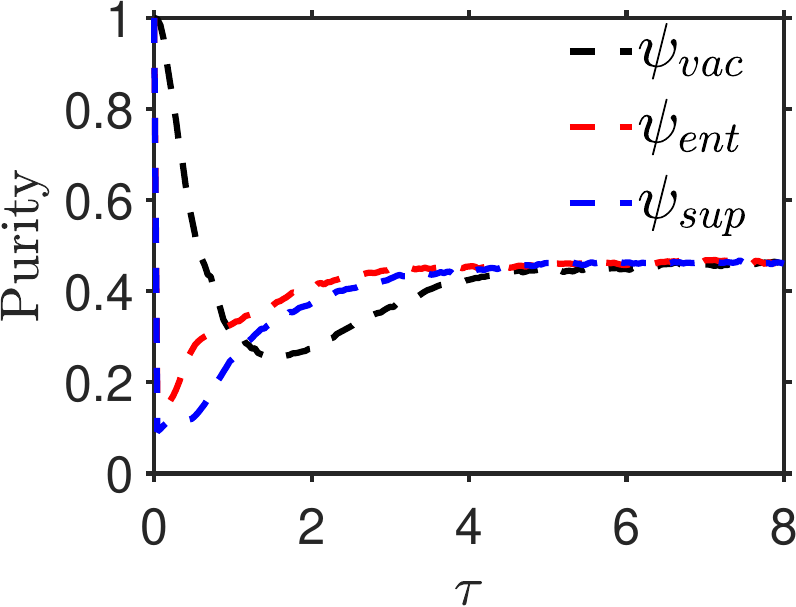}\includegraphics[width=0.4\columnwidth]{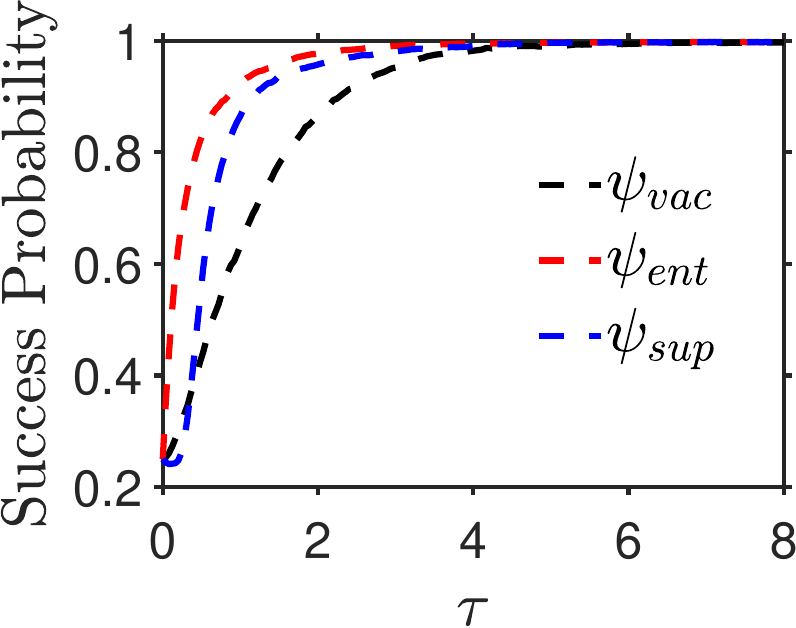}
\par\end{centering}
\caption{Left: Purity, and right: success probability dynamics for the $M=3$ case shown in Fig.~\ref{fig:Diag}, for different initial states over an extended time interval $\tau_{\max}=8$ with $\tau_{\mathrm{steps}}=200$. The quantum simulation reaches a steady state for the parameters $g=0.6$, $\xi_{0}=0.5$, $\lambda=2.4$, and $\alpha=3.87$, using a sufficiently large cutoff $N_{c}=20$, with errors less than $2\%$. Results are averaged over $10^{3}$ realisations. \label{fig:Purity_long}}
\end{figure}

 {  We next consider a low-nonlinearity regime ($0 \leq g \ll 1$) and examine how reduced nonlinearity affects the overall performance in terms of success probablity. The states in Fig. \ref{fig:Purity} are initialized with a coherent amplitude $\alpha = 3.87$, with parameters $\lambda = 2.4$ and $\xi_{0} = 0.5$. As shown in the top panels of Fig. \ref{fig:Purity}, the purity of highly quantum states initially decreases rapidly due to decoherence. At early times, the CIM explores a broad range of spin configurations. As the evolution proceeds and non-optimal configurations are progressively suppressed, the purity begins to recover and eventually saturates at the same level, for all initial states, if the same steady state with the same success probability is reached at long times, as shown in Fig. \ref{fig:Purity_long}.}

  { It is important to note that a cutoff of $N_c = 31$ is used in Fig. \ref{fig:Purity} for a simulation time of $t_{\max} = 0.5$. In the low-nonlinearity regime, this cutoff is sufficient within the considered time window, as the system has not yet reached its steady state. Extending the simulation to longer times would require a significantly larger cutoff; otherwise, truncation errors may affect the accuracy of the results.}

  {  A key insight from Fig.~\ref{fig:Purity} is that the performance of the vacuum state $\psi_{vac}$ is largely insensitive to changes in the nonlinearity parameter $g$, showing only a small decrease in success probability by the end of the simulation time, $\tau = 0.5$, as the nonlinearity parameter $g$ is decreased. In contrast, highly quantum states such as $\psi_{sup}$ and $\psi_{ent}$ exhibit improved performance in low-nonlinearity regimes, where the quantum states $\psi_{sup}$ and $\psi_{ent}$ reaches a higher success probablity at the end of the simulation time $\tau = 0.5$, compared to its high non-linearity counterparts. We attribute this behaviour to the fact that these states already encode components of the solution space, thereby reducing the reliance on tunneling to reach the optimal configuration. Consequently, the rate of purity degradation and its subsequent recovery appears to correlate with the overall success probability when the simulation is initialised with highly quantum states possessing features such as superposition and entanglement.}

\section{Conclusion}

We present fully quantum simulations of the Coherent Ising Machine
(CIM) in regimes where quantum effects are significant and cannot
be neglected. By employing number-state cut-offs well above the mean
photon number and leveraging quantum Monte Carlo techniques, we rigorously
examine the role of quantum phenomena in CIM performance.

Our study explores two categories of initial states: (1) the vacuum
state, and (2) a quantum superposition over possible Ising ground
states. We first investigate whether initializing the system with
a quantum superposition can enhance the probability of obtaining the
correct ground-state solution. By comparing these initial conditions,
we assess both success probabilities and the time required to reach
peak performance.

Two specific quantum superpositions are considered: a superposition
of outer products of cat states (Eq. \ref{eq:psi_sup}) and an $M$-partite
entangled state (Eq. \ref{eq:psi_ent}). Although our focus is on
these examples, the methods and conclusions can be extended to any
quantum state exhibiting similar properties.

Our results, as shown in Fig. \ref{fig:vary_alpha}, present strong evidence that
CIM experiments in with quantum superposition initial states could
achieve significant performance enhancements.

This conclusion that quantum enhancements are possible is further
supported by Figs \ref{fig:vary_alpha} and \ref{fig:Mean_Field},
where we compare the full quantum dynamics with classical mean-field
simulations. The results reveal a clear quantum advantage in parameter
regimes where classical false minima trapping cause classical CIM
computations to underperform. We attribute this improvement to quantum
tunneling, although other strategies might also give improvements
\cite{Zhou2025Frustration}.

In a second line of investigation, we initialize the system in the
vacuum state, as is standard in current experimental setups. We then
explore the effect of dynamically varying coupling strength over time, outlined in Sec. \ref{sec:dyn_str}.
As shown in Figures \ref{fig:a:-Ferro_vary_J},
certain fine-tuned strategies significantly increase the success probability.
This demonstrates that careful manipulation of system parameters can
lead to performance gains. However, such strategies may require tailored
implementations for different Ising instances. Other types of time-dependence
may give even better results, and clearly this approach has potential
for substantial improvements.

To understand the physical basis for these results, we further investigate
coherence loss through a purity analysis (Fig. \ref{fig:Purity}).
This shows that quantum superpositions undergo fast decoherence as
a result of dissipation. To address this, we experimented with initializing
the quantum state in a low nonlinearity regime, where the nonlinear
dissipation terms in Eq.\ref{eq:Mster_eq} are reduced. {We observe a performance
gain for initial quantum states with quantum properties, at lower dissipation
compared to the simulation starting with an initial vaccum state.}

This limitation reflects a fundamental aspect of quantum optical systems
with dissipative feedback coupling between modes, which leads to rapid
decoherence. This is true even when starting with an entangled pure
state. A non-dissipative coupling could mitigate this. Another limitation
of the results given here is that we focus on the internal state of
the CIM system, while it is the external current that is experimentally
accessible. There are techniques for simulating external homodyne
currents \cite{carmichael2009statistical}, including the homodyne
measurement shot noise, and the effects of this will be investigated
elsewhere \cite{Teh_homodyne_2026}.

In conclusion, our findings show that quantum effects can improve
success probabilities in CIMs, especially when classical solvers become
trapped in a false minimum. Our study should be viewed as a step toward
broader exploration of dynamical strategies and quantum state engineering
in future CIM designs.

\subsection*{Acknowledgments}
We gratefully acknowledge a grant from NTT Phi Laboratories. This
publication was made possible through the support of Grant 62843 from
the John Templeton Foundation. The opinions expressed in this publication
are those of the author(s) and do not necessarily reflect the views
of the John Templeton Foundation.



\section*{APPENDIX A: Quantum jump method}

Let the wave-function of the system at time $t$ be a normalized state
$|\psi(t)\rangle$. Using the MCWF method, the system's evolution
is governed by the Schr{\"o}dinger equation with a non-Hermitian effective
Hamiltonian:
\begin{align}
\mathcal{H}_{eff} & =\mathcal{H}_{sys}-\frac{i}{2}\underset{n}{\sum}C_{n}^{\dagger}C_{n},\label{eq:H_eff}
\end{align}
where the first term consists of the system\textquoteright s Hermitian
part of the Hamiltonian Eq. \ref{eq:Hamiltonian}, $\mathcal{H}_{sys}=i\frac{\gamma\lambda}{2}\sum\left(a_{j}^{\dagger2}-a_{j}^{2}\right)$,
and the second term represents non-Hermitian terms (often referred
to as Lindblad terms in other methods), including dissipation and
coupling, expressed by the jump operators $C_{n}$, which sums over
the dissipation terms.

We then calculate the wave-function $|\psi^{(1)}(t+\delta t)\rangle$
by evolving $|\psi(t)\rangle$ under the above non-Hermitian Hamiltonian
$\mathcal{H}_{eff}$:
\begin{align}
|\psi^{(1)}(t+\delta t)\rangle & =\left(1-i\delta t\mathcal{H}_{eff}\right)|\psi(t)\rangle.
\end{align}
Here $\delta t$ must be sufficiently small, allowing us to ignore
higher-order terms $\delta t^{2}$ and above. Next, we evaluate the
square of the norm:
\begin{align}
|{\rm norm}|^{2} & =\langle\psi^{(1)}(t+\delta t)|\psi^{(1)}(t+\delta t)\rangle\nonumber \\
 & =\langle\psi(t)|\left(1+i\delta t\mathcal{H}_{eff}^{\dagger}\right)\left(1-i\delta t\mathcal{H}_{eff}\right)|\psi(t)\rangle\nonumber \\
 & =1-\delta p.\label{eq:norm2}
\end{align}

We use a slightly modified version of the original algorithm proposed
by Molmer et al. \cite{Molmer_Qjump}, where the time integration
differs depending on the software package used. We developed two separate
codes to allow cross-checking. One is in MATLAB using the xSPDE4 library
\cite{kiesewetter2016xspde,Drummond2025xspde4} which uses 4th-order
Runge-Kutta method for integration , and another in Python utilizing
the QuTiP library \cite{qutip1,qutip2}, which employs an adaptive
Runge--Kutta method. The input codes for the numerical simulations
can be found in an online archive \cite{DVN/Z7GSKY_2025}.

Since the evolution occurs under a non-Hermitian Hamiltonian $H_{eff}$,
the norm of the wave-function decreases by $\delta p$. We integrate
the system  until a certain threshold, determined by a uniformly
distributed random number $\epsilon$, is reached. Once this threshold
is met, a quantum jump occurs, and the system evolves according to
one of the collapse operators, projecting the state $|\psi(t)\rangle$using
the collapse operator $C_{n}$:
\begin{align}
|\psi^{(1)}(t+\delta t)\rangle & =\frac{C_{n}|\psi(t)\rangle}{\langle\psi(t)|C_{n}^{\dagger}C_{n}|\psi(t)\rangle^{1/2}}.
\end{align}

The collapse operators $C_{n}$ correspond to the dissipative terms
in our master equation, given by last three terms in Eq. \ref{eq:Mster_eq}.
The associated Lindblad operators are $a_{i}$, $a_{i}^{2}$ and $a_{i}-\frac{J_{ij}}{|J_{ij}|}a_{j}$,
where $i$ and $j$ denote different modes, so that $i,j=1,2,...M$.

When multiple collapse operators are present, another random number
$r$ is generated to select which collapse operator causes the jump.
The jump occurs when the following inequality is satisfied: 
\begin{align}
\overset{n}{\underset{i=1}{\sum}}P_{i} & \geq r,
\end{align}
where $n$ is the smallest integer that satisfies the inequality,
and $P_{i}=\langle\psi(t)|C_{i}^{\dagger}C_{i}|\psi(t)\rangle/\delta p$.

At a given time $t$, the  approximate density matrix $\rho$ for $N_s$ trajectories $ |\psi_i(t)\rangle$ in the ensemble is then given by:
\begin{align}
\rho(t) & = \sum_{i=1}^{N_s}  |\psi_i(t)\rangle \langle\psi_i(t)|.
\end{align}

\section*{APPENDIX B}

It is important to understand the potential errors in any numerical
calculation. For a stochastic calculation there are both time-step
and sampling errors, which we monitor as follows.

\subsection*{Time-Step Error Calculation}

We next estimate the time-step error, which arises due to the finite
time discretization used in the numerical simulation. This error is
evaluated by comparing the results obtained with two different time
step sizes: a coarse time step and a refined time step that is half
the size of the coarse one.

To ensure consistency in numerical simulations, both simulations use
the same random number sequence. For each coarse step, two fine steps
are taken to reach the same time point. This allows for an accurate
comparison between the results at the same physical times.

At each coarse time point $t_{i}$ , we define the deviation between
the fine and coarse results as:
\begin{align}
\epsilon_{c}(t_{i}) & =O_{fine}(t_{i})-O_{coarse}(t_{i}),
\end{align}
where $O_{fine}(t_{i})$ and $O_{coarse}(t_{i})$ are the observable
values obtained using the fine and coarse time steps, respectively.
The time-step error is then estimated by computing the root mean square
(RMS) of these deviations over all time points:
\begin{align}
\epsilon_{t_{c}} & =\sqrt{\frac{1}{N_{t}}\underset{i=1}{\overset{N_{t}}{\sum}}|\epsilon_{c}(t_{i})|^{2}}.
\end{align}

Finally, to express this error in a relative form, we define the normalized
relative time-step error as:
\begin{equation}
\epsilon_{Ns}=\frac{\epsilon_{t_{c}}}{\max(|O|)},
\end{equation}
where $\max(O)$ is the maximum absolute value of the observable across
all time points.

\subsection*{Sampling Error Calculation}

The sampling error is estimated using a two-level sub-ensemble averaging
method so that the central limit theorem is applicable. The total
ensemble size is defined as $N_{s}=N_{s}^{(1)}\times N_{s}^{(2)}$,
where $N_{s}^{(1)}$ denotes the number of trajectories in each sub-ensemble
(serial sampling), and $N_{s}^{(2)}$ represents the number of sub-ensembles
(parallel sampling). In our simulations, $N_{s}^{(1)}$ ranges from
$100$ to $500$, and we use $N_{s}^{(2)}=10$.

Let $O_{i}=\langle\psi_{i}|\hat{E}|\psi_{i}\rangle$ be the observable
evaluated from each trajectory $|\psi_{i}\rangle$, and $\hat{E}$
is a Hermitian operator corresponding to the quantity being measured.
First, we compute the mean over the lower-level ensemble $\bar{O}_{(j)}$,
which is the average over the lower ensembles $N_{s}^{(1)}$ as
\begin{align}
\bar{O}_{(j)}(t_{k}) & =\frac{1}{N_{s}^{(1)}}\overset{N_{s}^{(1)}}{\underset{i=1}{\sum}}O_{(i)}(t_{k}),
\end{align}
at each time point $t_{k}$, where $k=1,2,...N_{t}$, for $N_{t}$
number of time points. Next, we evaluate overall sample mean $\bar{O}$
by taking the sample averages $\bar{O}_{(j)}$ averaged over the higher
ensembles $N_{s}^{(2)}$ as
\begin{align}
\bar{O}(t_{k}) & =\frac{1}{N_{s}^{(2)}}\overset{N_{s}^{(2)}}{\underset{j=1}{\sum}}\bar{O}_{(j)}(t_{k}).
\end{align}

The variance $\sigma^{2}$ of this data is used to estimate a standard
deviation $\sigma$ in the mean, giving
\begin{align}
\sigma & (t_{k})=\sqrt{\overset{N_{s}^{(2)}}{\underset{j=1}{\sum}}\frac{\left|\bar{O}_{(j)}(t_{k})-\bar{O}(t_{k})\right|^{2}}{N_{s}^{(2)}-1}}.
\end{align}

According to the central limit theorem, the distribution of the sample
means approaches a normal distribution as the sample size increases.
Hence, the standard error of the mean at each time point is:
\begin{align}
\sigma_{s}(t_{k}) & =\frac{\sigma(t_{k})}{\sqrt{N_{s}^{(2)}}}.
\end{align}

The overall sampling error is then computed as the root-mean-square
of the standard error over all time points $N_{t}$:
\begin{align}
\epsilon_{s} & =\sqrt{\frac{1}{N_{t}}\overset{N_{t}}{\underset{n=1}{\sum}}\sigma_{s}^{2}(t_{k})},
\end{align}
and finally, we define the normalized relative sampling error as a
dimensionless quantity:
\begin{align}
\epsilon_{Ns}= & \frac{\epsilon_{s}}{\max(O|)}.
\end{align}
Here $\max|O|$ is the maximum absolute value of the observable across
all time points.

\section*{References}
\bibliographystyle{iopart-num}

\providecommand{\newblock}{}

\end{document}